\documentclass{aa}  

\usepackage{graphicx}
\usepackage{aalongtable,lscape}
\usepackage{txfonts}
\usepackage{natbib}
\bibpunct{(}{)}{;}{a}{}{,} % to follow the A&A style.

\begin{document}

%\title{Optical imaging and spectroscopy of candidate GPS radio sources\thanks{Based on observations collected at the European Southern Observatory, Chile (ESO Programmes 64.P-0482(A) and 68.B-0044(A)).}}

\title{GPS radio sources: new optical observations and an updated master list}
%\thanks{Based on observations made with ESO Telescopes at Paranal Observatory under Programmes 64.P--0482(A) and 68.B--0044(A)).}}

\author{
A. Labiano\inst{1,}\inst{2}
\and
P. D. Barthel\inst{1}
\and
C. P. O'Dea\inst{3}
\and
W. H. de Vries\inst{4}
\and
I. P\'erez\inst{1}
\and
S.A. Baum\inst{5}
}
\offprints{Alvaro Labiano:\\ {\tt labiano@damir.iem.csic.es} \smallskip}

\institute{
Kapteyn Astronomical Institute, Groningen, 9700 AV, The Netherlands %\\
\and
Departamento de Astrof\'isica Molecular e Infrarroja, Instituto de Estructura de la Materia (CSIC), Madrid, Spain
\and
Department of Physics, Rochester Institute of Technology, Rochester, NY, 14623, USA%\\
\and
Lawrence Livermore National Laboratory, Livermore CA, 94550, USA%\\
\and
Center for Imaging Science, Rochester Institute of Technology,  Rochester, NY 14623. USA %\\
}

\date{  }

\abstract
{}
{ Identify optical counterparts, address uncertain identifications and measure previously unknown redshifts of the host galaxies of candidate GPS radio sources, and study their stellar populations.}
{Long slit spectroscopy and deep optical imaging in the $B$, $V$ and $R$ bands, obtained with the Very Large Telescope.}
{We obtain new redshifts for \object{B0316+161}, \object{B0407--658}, \object{B0904+039}, 
\object{B1433--040}, and identify the optical counterparts of \object{B0008--421}  and \object{B0742+103}. 
We confirm the previous identification for \object{B0316+161}, \object{B0407--658}, \object{B0554--026}, and 
\object{B0904+039},  and find that the 
previous identification for \object{B0914+114} is incorrect. Using updated published 
radio spectral information {we classify as non GPS}  the following sources: \object{B0407--658}, \object{B0437--454}, 
\object{B1648+015}. \\ The optical colors of typical GPS sources are  
consistent with single instantaneous burst stellar population
models but do not yield useful information on age or metallicity. A new master list of GPS sources is presented.}
{} 
 

\maketitle
%
%________________________________________________________________

\section{Introduction}
{GigaHertz} Peaked Spectrum (GPS) radio sources are fairly common -- \citet{O'Dea98} lists a 10\% fraction 
in high frequency selected catalogs.  In the current paradigm, the extended radio 
galaxies and quasars are unified with the compact, core-dominated quasars and  BL Lac-objects 
through the combined effects of radio jet orientation and anisotropic obscuration 
\citep[e.g.,][]{Urry95}.  These objects are considered to be mature, well developed radio 
sources.  It is likely that GPS objects are young radio sources that will evolve into the 
10 -- 100 kpc scale objects. Studies of radio galaxy evolution  
suggest that the GPS sources can evolve into the larger scale radio
galaxies, dimming in radio luminosity as they expand  
\citep[e.g.,][]{Fanti95, Begelman96, Readhead96b,Young97, Kaiser97a, Kaiser97b, O'Dea97, Snellen00}. 
Multi-color optical as well as near-IR imaging \citep[e.g.,][]{O'Dea96, Snellen96, 
Vries00} have shown that host galaxy colors of nearby GPS objects are indeed consistent with 
non- or passively evolving ellipticals, with absolute magnitudes comparable to brightest 
cluster members, similar to the hosts of intermediate sized and large radio source classes. 
Determination of the rest-frame broad-band colors (which requires redshifts) in connection 
with stellar population synthesis modeling has proven essential for these 
investigations \citep[e.g.,][]{Vries00b}.

\citet{O'Dea91} presented a list of candidate GPS radio sources. About half of those sources had 
unknown redshifts and several lacked an optical counterpart. The sample has been updated with 
new identifications and redshifts since then \citep[e.g.,][]{Vries97b}. However, the optically faint 
part of the sample requires the use of 8 m class telescopes. Our goal in this paper is to increase the number 
of optical identifications, measure new redshifts and remove non GPS sources 
%present in the list (with the currently available radio spectral data). 
%{\bf from current samples, once all the information available in the literature allow to derive more accurate radio spectra. }
{from the current samples, using updated radio data from the literature.}
Using the ESO VLT, we have therefore observed 8 unidentified and 5 tentatively
identified objects from the  master list of \citet{O'Dea91}.

We here present these observations of sources with unknown or uncertain 
 redshifts or optical counterparts, 
and increase the number of confirmed GPS sources (up to 74, out of 95 candidates) of the complete     
\citet{O'Dea91} list. We give new host identifications (down to $R \sim$ 25) and obtain 
several new redshifts. 
Using these, as well as literature research and our earlier data, we update the \citet{O'Dea91} master list. This updated
   master list is presented in a concluding Table.
 We use H$_0$=71, $\Omega_M = 0.27 ,  \Omega_\Lambda = 0.73$ 
\citep{Spergel03} throughout the paper.

\section{Observations and data reduction}

The sample was observed during two nights (January 30 to 31, 2000, and December 16 to 17, 2001) using VLT's FORS1/UT1 and FORS2/UT4 \citep{Appenzeller98}. We obtained long slit spectroscopy using grism 150I with order separators OG590 and GG375, obtaining a 230 \AA/mm dispersion (5.52~\AA/pixel) and covering wavelengths 6000--11000~\AA\ (OG590) and 3850--7500\AA\ (GG375). For the imaging we used a {\it R}-Johnson-Cousins filter for the January 2000 run. For the December run, we used the {\it B}, {\it V}- Johnson-Cousins and {\it R}-Special filters\footnote{The {\it R}-Special filter on FORS2 is a Johnson-Cousins filter with a slightly shortened red end to avoid sky emission lines.}. We covered a 6.8 arcmin field with seeing ranging from 0.5\arcsec\ to 0.7\arcsec\ for the January run and 0.5\arcsec\ to 1.1\arcsec\ for the December run. The pixel scale of FORS is 0.25\arcsec/pixel. Table \ref{exptimes} lists the exposure times.
%The spatial resolution of FORS is $\sim0.2\arcsec$. Wim me dice que lo quite y ponga pixel scale.

Standard data reduction was performed using IRAF routines. All the spectra were corrected for bias, flat-field and sky subtracted. Wavelength calibration was done using internal arc lamps. The flux calibration and removal of atmospheric lines were performed using the spectrophotometric standards GD50 and GD108. %\citep[e.g.,][]{Nitschelm88} para ua referencia con lineas de absorcion atmosferica.

\begin{table}[t]
\caption{Source list and exposure times (in seconds).}
\label{exptimes}
\begin{minipage}{\columnwidth}
\centering
%\resizebox{\textwidth}{!}{
\begin{tabular}{lccccccc}
\hline
\hline
%             & \multicolumn{2}{c}{Radio Position} & & \multicolumn{2}{c}{Exposure time (s)}\\
%\cline{2-3}
%\cline{5-6}
Name & Spectroscopy & $B$ & $V$ & $R$ \\
\hline 
\object{B0008--421}    & 1800 & -- & -- & 600\\
\object{B0316+161}    & 3000 & 300 & 600 & --\\
\object{B0407--658}    & 1500 & 600 & 300 & 300 \\
\object{B0437--454}    & 1800 & -- & -- & --  \\
\object{B0554--026}    & -- & 300 & 300 & 300  \\
\object{B0742+103}    & 3000 & -- & -- & 600\\ 
\object{B0904+039}    & 3000 & 600 & 300 & -- \\
\object{B0914+114}    & 1200$^a$ & 600 & 480 & --  \\
\object{B1045+019}           & 1800 & -- & -- & --\\
\object{PMN J1300--1059}$^b$  & 900 & -- & -- & -- \\
\object{B1433--040}            & 1200 & -- & -- & -- \\
\object{B1601--222}    & 1200 & -- & -- & -- \\
\object{B1648+015}    & 1350 & -- & -- & -- \\
\hline
\end{tabular}
%}
\end{minipage}
\\ \\
$^a$ Unrelated galaxy. See notes on individual source for details.\\
$^b$  \object{PMN J1300--1059} was chosen from the NVSS and WISH surveys \citep{Condon98, Breuck02}.

\end{table}

The calibration of the imaging data was different for the January and December runs. Both runs were corrected for bias and flat-field. The January observations were done during a photometric night. We took flat field images and observed the \citet{Landolt92}  standard fields PG1323--086 and SA 95. No useful standard fields or flat fields were observed for the --non photometric-- December run. To correct from flat-field  these images, we averaged all the observations in each filter separately to create {\it artificial} flat-field images, which we used for the correction. As standard stars, we chose unsaturated field stars with data in the Second Guide Star Catalogue \citep[GSC2.2,][]{McLean98}, {about} 40 stars in total.  The GSC stars have available magnitudes in the photographic F and J bands. The transformation to the Johnson-Cousins filters was performed following \citet{Kent85}. The January images were taken only in $R$-band. For the December run, most of the standards lacked color information to perform (first order) color coefficient corrections to our apparent magnitudes. We could only fit the zero point magnitudes in each band. The errors on the zero point are: 0.042 for the January observations ($R$-band) and 0.38, 0.17, 0.10 for the December observations ($R$,$B$,$V$- bands respectively).

\section{Results}

\subsection{Identifications}

The astrometry was performed using the GSC2.2 catalog as reference. It was possible to make accurate (usually with $1\sigma$ error $<$ 0.5 arcsec) positional determinations for the candidate optical counterparts. {The radio positions are from the VLA calibrator manual (\object{B0008-421}, \object{B0316+161}, \object{B0742+103}) and NED (\object{B0407--658}, \object{B0554--026}, \object{B0904+039}, \object{B0914+114})}. As in \citet{Vries95, Vries00b}, we use the {\it likelihood ratio} defined by \citet{Ruiter77}:

\begin{equation}
\hfil  R_1 = \sqrt{ \frac{\Delta\alpha^2}{\sigma^2_\alpha} + \frac{\Delta\delta^2}{\sigma^2_\delta}}  
\end{equation}
where $\Delta\alpha$ and $\Delta\delta$ are the measured offsets in RA and Dec between the optical 
and radio positions 
%(Figure \ref{Deltas})
, $\sigma^2_\alpha$ and $ \sigma^2_\delta$ are the sums of the squared 
1$\sigma$ errors in the optical and radio positions.   
The probability that a true optical  counterpart has an R$_1$ value larger than some 
R$_0$ is given by P($R_1>R_0$) = e$^{-0.5R^2_0}$. An R$_1$ value less than three indicates 
a probability of less than 1\% of a false identification (assuming that the optical counterpart 
is the object closest to the radio position).  Most of 
our identifications have $R_1$ values smaller than 3 and are likely to be correct. The optical 
position used in the calculation of $R_1$, and listed in Table \ref{Positions}, corresponds 
to the center of the source, fitted with IRAF task Ellipse. The identification results are 
listed in Table \ref{Positions} and finding charts for the sources are presented in Figure \ref{maps}.  %the brightest pixel of the optical source

%\input{CHAPTER3/TABLES/Positions.tex}

%%%%% TABLE 1 %%%%%
\begin{table*}[t]
\caption{Positions of the optical counterparts.} 
\label{Positions}
\begin{minipage}{2\columnwidth}
\centering
%\resizebox{\textwidth}{!}{
\begin{tabular}{ccccccl}
\hline
\hline
  &  \multicolumn{2}{c}{Radio (J2000)} & \multicolumn{2}{c}{Optical (J2000)} \\
\cline{2-3}  
\cline{4-5}
Source   &  RA & Dec  & RA & Dec & $R_1$ & Magnitude\\
\hline 
\object{B0008--421}    & 00:10:52.52        & --41:53:10.8 & 00:10:52.53        & --41:53:10.6 & 2.1 & $R$ 24.3$\pm$0.3$\pm$0.04 \\
\object{B0316+161}    & 03:18:57.80        & +16:28:32.7 & 03:18:57.82        & +16:28:32.9 & 2.0 & $B$ 23.0$\pm$0.1$\pm$0.2  \\
\object{B0316+161}    & " & "                                                 & 03:18:57.82        & +16:28:32.6 & 2.0        & $V$ 23.4$\pm$0.1$\pm$0.1  \\
\object{B0407--658}    & 04:08:20.38        &--65:45:09.1 & 04:08:20.37        & --65:45:09.0 & 1.3 & $B$ 22.5$\pm$0.04$\pm$0.2  \\
\object{B0407--658}    &"&"                                                & 04:08:20.41        & --65:45:08.6 & 3.4 & $V$ 21.4$\pm$0.03$\pm$0.1  \\
\object{B0407--658}    &"&"                                                & 04:08:20.41        & --65:45:09.4 & 3.4 & $R$ 20.2$\pm$0.06$\pm$0.4  \\
\object{B0554--026}    & 05:56:52.62        &--02:41:05.5 & 05:56:52.61        & --02:41:05.3 & 2.6 & $B$ 18.3$\pm$0.01$\pm$0.2 \\
\object{B0554--026}    &"&"                                                & 05:56:52.59        & --02:41:05.5 & 3.3 & $V$ 17.5$\pm$0.02$\pm$0.1 \\
\object{B0554--026}    &"&"                                                & 05:56:52.62        & --02:41:05.5 & 3.2 & $R$ 16.4$\pm$0.01$\pm$0.4 \\
\object{B0742+103}    & 07:45:33.06        &+10:11:12.7 & 07:45:33.06        & +10:11:12.5 & 1.9 & $B$ 24.0$\pm$0.1$\pm$0.2 \\
\object{B0742+103}    &"&"                                                & 07:45:33.06        & +10:11:12.3 & 1.9 & $V$ 23.8$\pm$0.2$\pm$0.1 \\
\object{B0742+103}    &"&"                                                & 07:45:32.97   & +10:11:12.8 & 3.9 & $R$ 23.1$\pm$0.1$\pm$0.04 \\
\object{B0904+039}    & 09:06:41.05        &+03:42:41.5 & 09:06:41.03        & +03:42:42.0 & 3.3 & $V$ 24.9$\pm$0.3$\pm$0.1 \\
\object{B0914+114}    & 09:17:16.39        & +11:13:36.5 & 09:17:16.53        & +11:13:31.4 & 52.6 & $B$ 21.3$\pm$0.04$\pm$0.2 \\
\object{B0914+114}    &"&"                                                & 09:17:16.54        & +11:13:32.3 & 44.4 & $V$ 19.9$\pm$0.01$\pm$0.1 \\

\hline
\end{tabular}
%}
\end{minipage}
\\ \\
The radio positions are from the VLA calibrator manual (\object{B0008-421}, \object{B0316+161}, \object{B0742+103}) and NED (\object{B0407--658}, \object{B0554--026}, \object{B0904+039}, \object{B0914+114}). The typical uncertainties in the radio positions are $~0.01\arcsec$ in Declination and $~0.001$s in Right Ascension. The $R_1$ factor is the likelihood ratio (see text for details). The optical coordinates of \object{B0914+114} correspond to the galaxy previously -- and incorrectly -- identified as the counterpart (see notes on individual source). The errors in the magnitudes are divided in photon noise (first error) and calibration (second error). The magnitudes are corrected for Galactic extinction, following \citet{Schlegel98}. \object{B0316+161} magnitudes may be affected by close-by ($\sim$3\arcsec) objects. 
\end{table*}

\subsection{Magnitudes}
\label{subsec:mags}

The magnitudes were extracted performing aperture photometry. We used apertures large enough to include all light from the object (typically a diameter of $\sim10$ pixels) but minimizing the contribution from sky. {The magnitudes are corrected for Galactic extinction, following \citet{Schlegel98}}. 

\object{B0008--421} and \object{B0742+103} were observed in the {\it R}-Johnson-Cousins filter in the January 2000 --photometric-- night.  The photon noise increases the error in magnitude to 0.3 and  0.1 for \object{B0008--421} (with lower signal to noise) and \object{B0742+103} respectively (Table \ref{Positions}). %The error of the zero point fit is 0.042 for both sources ({\bf WE SAID THIS BEFORE--REMOVE IT FROM HERE?}) 

%The low signal to noise in the image of \object{B0008--421} increases the total error to 0.3 while 
%complicates the fit of the source, increasing the total error to 0.3.
%T Al tener peor S/N, phot se lia al ajustar y calcular la magnitud.
%The total error in magnitude is larger for \object{B0008--421}, due to the faintness of the source.

The rest of the images were taken in the --non photometric-- December 2001 run. As judged from internal consistency of field stars with known magnitudes, our photometric accuracy varies between 0.1 and 0.4 magnitudes. The main source of error for this run comes from the uncertainties in the magnitudes of the stars used in the calibration: 0.17, 0.10 and 0.38 for $B$,$V$ and $R$ bands respectively. A conservative ($5\sigma$) detection limit is magnitude $\sim$25.5 for $R$, $V$-bands and $\sim$26 for $B$-band, for all sources.

\begin{figure*}[tbp]
\centering
\includegraphics[width=2\columnwidth]{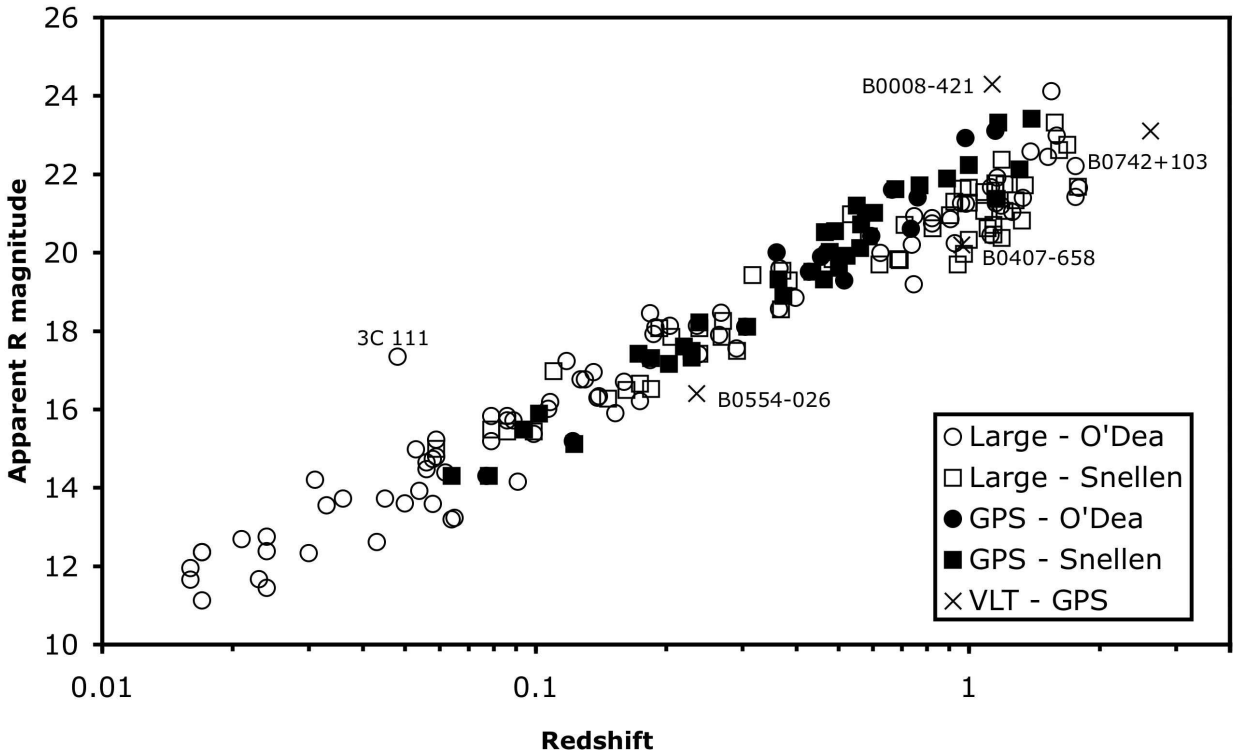}
\caption{$R$-band Hubble diagram for GPS and large radio galaxies. Data from \citet{O'Dea96}, \citet{Snellen02} and our VLT measurements (marked with X). Redshifts of \object{B0742+103} and \object{B0554--026} from \citet{Best03} and \citet{Vries00b}.  \object{3C~111} is obscured by the Galactic dark cloud complex Taurus B \citep[e.g.,][ and references therein]{Sguera05}. }

\label{Hubble}
\end{figure*}

Comparison with the Hubble diagram \citep[][ and Figure \ref{Hubble}]{O'Dea96, Snellen02} show that the R-magnitudes we measure are consistent with previous observations of GPS hosts. The V-magnitudes of the new optical counterparts are consistent with those found by \citet{Serego94} for hosts of radio sources, although our sources tend to be slightly fainter. We have not found published B magnitudes for comparison.
%We measure V-magnitudes consistent with those from \citet{Serego94} although ours tend to be slightly fainter. We have not found published B magnitudes for comparison.

\subsection{Spectra and redshifts}

We took spectra of 16 sources to measure their redshifts. Table \ref{Lines} and Figure \ref{spec} show those spectra where emission features were found. We measure redshifts for \object{B0316+161}, \object{B0407--658}, \object{B0904+039}, \object{B1433--040} (with a conservative error of $\pm$0.001) and (the incorrect ID for)
 \object{B0914+114}. Some sources show possible emission lines but their redshifts are uncertain (see notes on each individual object below): \object{B0008--421}, \object{PMN J1300--1059},  and \object{B1045+019}. The spectra were too noisy to find emission features for \object{B0437--454}, \object{B0742+103}, \object{B1601--222} and \object{B1648+015}.

%\input{CHAPTER3/TABLES/Lines.tex}

%%%%% TABLE 2 %%%%%

\begin{table*}[t]
\caption{Line identifications and redshifts.}
\label{Lines}
\begin{minipage}{2\columnwidth}
\centering
%\resizebox{\textwidth}{!}{
\begin{tabular}{ccccccccc}
\hline
\hline
Name & ID  & Line & Wavelength(\AA) & Redshift & Mean  &Flux &FWHM& Log Power 5GHz\\
\hline 
\object{B0008--421}  & G  &     [\ion{Ne}{v}] 3425 ? & 7294 & 1.130 & 1.130* & 0.3&29&26.7\\
                  &   &       [\ion{O}{ii}] 3727 ? & 7940 & 1.130& &0.7&39 \\
%\object{B0008--421}  & G  &     H$_\beta$ 4861 ? & 7084 & 0.457 & 0.457* & 0.2&13&26.7\\
%                  &   &       [\ion{O}{iii}] 5007 ? & 7294& 0.457&&0.3&29 \\
\hline
\object{B0316+161}   &  G   & [\ion{O}{ii}] 3727 & 7102 & 0.906 &  0.907 &5.4&21&27.6\\
                    &   &   [\ion{Ne}{iii}] 3869 & 7381& 0.908 & &1.8&32\\
%                  &   &   [\ion{Ne}{iii}] 3967 & 7565& 0.907 & \\
                  &   &     H$_\beta$ 4861 & 9277 & 0.908 & &4.7 & 49\\
                  &   &     [\ion{O}{iii}] 4959 & 9464& 0.908& &8.3 & 35\\
                  &   &       [\ion{O}{iii}] 5007 & 9544& 0.906&&33 &51\\
\hline
\object{B0407--658}  & G  & [\ion{Ne}{v}] 3425 & 6721 & 0.962 & 0.962 & 11&25 &27.7\\ 
                  &   & [\ion{O}{ii}] 3727 & 7314 & 0.962  &   &6.9&29\\
                    &   &   [\ion{Ne}{iii}] 3869 & 7595& 0.963 & &2.0&44\\
                  &   &   [\ion{Ne}{iii}] 3967 & 7787& 0.963 & &0.7&29\\
%                  &   &     H$_\delta$ 4861 & 9538 & 0.962 & \\
                  &   &     H$_\gamma$ 4102 & 8048 & 0.962 && 0.3& 18\\
                  &   &     H$_\beta$ 4861 & 9538 & 0.962 && 4.4&52\\
                  &   &     [\ion{O}{iii}] 4959 & 9731& 0.962& &43&51\\
                  &   &       [\ion{O}{iii}] 5007 & 9823& 0.962& &16&54\\
\hline

%\object{B0742+103}   &  & [\ion{O}{ii}] 3727 & 8477 & 1.274 & 1.275*  \\
%                    &   &       [\ion{Ne}{iii}] 3869 & 8802 & 1.275 & \\

\object{B0904+039}   & G & [\ion{O}{ii}] 3727 & 6822 & 0.830 & 0.830&  1.9&17& 24.9\\
                  &   &       [\ion{O}{iii}] 4959 & 9078& 0.831&& 1.8&26\\
                  &   &      [\ion{O}{iii}] 5007 & 9162& 0.830&& 4.4&26\\
\hline
\object{B0914+114} & G & Break 4000 & 4711 & 0.178 & 0.178 &  --&--& --\\
                   &  &  H$\beta$ 4861 & 5690?  & 0.171? & $\pm$0.005&  --&--\\
                   &  &  Mg{\it b} 5200 & 6115 & 0.183& &  --&--\\
                   &  &  [\ion{N}{ii}] 6548 & 7667 & 0.171& &  --&--\\
                &  & H$\alpha$ 6563 & 7772 & 0.184 &  &--&--\\
                   &  &  [\ion{S}{ii}] 6716 & 7952 & 0.184& &  --&--\\
\hline            
\object{B1045+019}  &  &  [\ion{O}{ii}] 3727 & 6297 & 0.689 &  0.689*&0.4&12&26.6 \\
                  &   &      H$_\gamma$ 4102 & 7327 & 0.688 &&   0.3&12\\
                  &   &      [\ion{He}{ii}] 4686 & 7918& 0.689&& 0.3&29\\
\hline
\object{PMN J1300--1059}   &Q?   &  \ion{Mg}{ii} 2799  & 6385 & 1.283 & 1.283*&10&125&--  \\
\hline
\object{B1433--040}   & Q &  [\ion{O}{ii}] 3727 & 6694 & 0.796 &  0.795& 17&19&26.3\\
                    &   &     [\ion{Ne}{iii}] 3869 & 6949& 0.796 && 32 & 54\\
                  &   &     [\ion{Ne}{iii}] 3967 & 7122& 0.795 && 10&21\\
                  &   &      H$_\delta$ 4102 & 7355  & 0.794  &&  23&84\\
                  &   &      H$_\gamma$ 4341 & 7790 & 0.795 &&  76&113 \\
                  &   &      H$_\beta$ 4861 & 8724 & 0.795 && 347&192\\
                  &   &      [\ion{O}{iii}] 4959 & 8900& 0.795&& 63&30\\
                  &   &      [\ion{O}{iii}] 5007 & 8987& 0.795&& 121&27\\
\hline
\end{tabular}
%}
\end{minipage}
\\ \\
For \object{B1433--040}, the redshift is measured with the narrow emission profiles. The * means the redshift estimation is uncertain for that source, see text for details. The FWHM listed is the observed FWHM corrected for the FWHM of the instrumental spectrum. Fluxes in 10$^{-16}$ erg/cm$^2$/s. A conservative error for {\it z} is $\pm$ 0.001. The \object{B0914+114} results correspond to the unrelated galaxy. It shows emission and absorption lines mixed so the centers are more uncertain than in the other sources (the 1$\sigma$ ~error in {\it z} is listed below the mean value) and line properties will depend on the stellar population model. The last column lists the radio power in W/Hz of the source for the redshifts listed. The 5~GHz fluxes are from \citet{O'Dea98} and \citet{Wright90}.
\end{table*}
%%%%%%%%%%%%%%%

\subsection{Stellar population synthesis models}

GPS radio sources are found to live in passively evolving elliptical galaxies, and have colors consistent with those predicted by stellar synthesis models \citep[e.g.,][]{O'Dea96, Vries98b, Snellen02}. Most of our objects are identified with {\it galaxies} and not quasars. Therefore, most of the optical radiation can be associated with stars in the host. 
%Furthermore, previous work \citep{O'Dea96, Snellen96} find that the $R$ (and $K$) band magnitudes correlate with redshift for GPS and 3CR galaxies as predicted by stellar synthesis population models.
%, indicating that the emission is dominated by the host stellar population.
%follow the predictions of sspms...!!!!!!!!! 

%Most of our objects are radio galaxies so we expect the contribution from the AGN emission to be minimal: most of the radiation would be associated with stars in the host. 
%De Wim 1998:    
%In fact R- and K-band magnitudes for these radio galaxies have been found to correlate with redshift (the Hubble diagram, e.g., O'Dea et al. 1996; Snellen et al. 1996), indicating that the emission is dominated by the host stellar population.  Also, none of the sources were found to be dominated by a point source in the near-IR (Paper I), which is expected for an obscured AGN, and the surface brightness profiles all indicated the sources were resolved. Therefore, we feel confident that the R-K (and J-K) colors are due to the underlying host stellar population, and that comparison between observations and stellar synthesis models is meaningful. 

%Idea?
%This is how understand it (and what I tried to say):
%O'Dea96 and Snellen96 find that the $R$ (and $K$) band magnitudes correlate 
%with redshift for GPS and 3CR galaxies as predicted by stellar synthesis 
%population models. So the main contribution should come from star light.
%I got the idea from de Vries et al. 1998,AJ, 503, 156 (Section 4) and then 
%checked the original papers.

We have compared our redshift and magnitude measurements with the \citet{Bruzual03} stellar population synthesis models. The models have been run using the Chabrier \citep{Chabrier03} initial mass function, metallicities Z=0.008, Z=0.02, Z=0.05 and ages (time since the initial starburst) ranging from 5 to 12.5 Gyr. Different star formation histories have been used: instantaneous burst, exponentially declining, and constant star formation. We fitted colors $B-V$ and $V-R$ for the sources where we had color information.

%We fitted colors $B-V$ and $V-R$ for those sources where the color information was available, and magnitudes when color information was not available. The hosts of GPS/CSS are usually massive elliptical galaxies \citep[e.g.,][]{Snellen99} so we used masses of 10$^{11}$ and 10$^{12}$ M$_\odot$ \citep[e.g.,][]{Rocca04} to fit those sources without color information.

Constant and exponentially declining star formation models seem to be ruled out by our data. The data points and best models disagree by roughly  0.5 to 1 magnitude .
%for those sources with color information. 
% 2 magnitudes for sources without color information and
\begin{figure}[h]
\centering
\includegraphics[width=0.7\columnwidth]{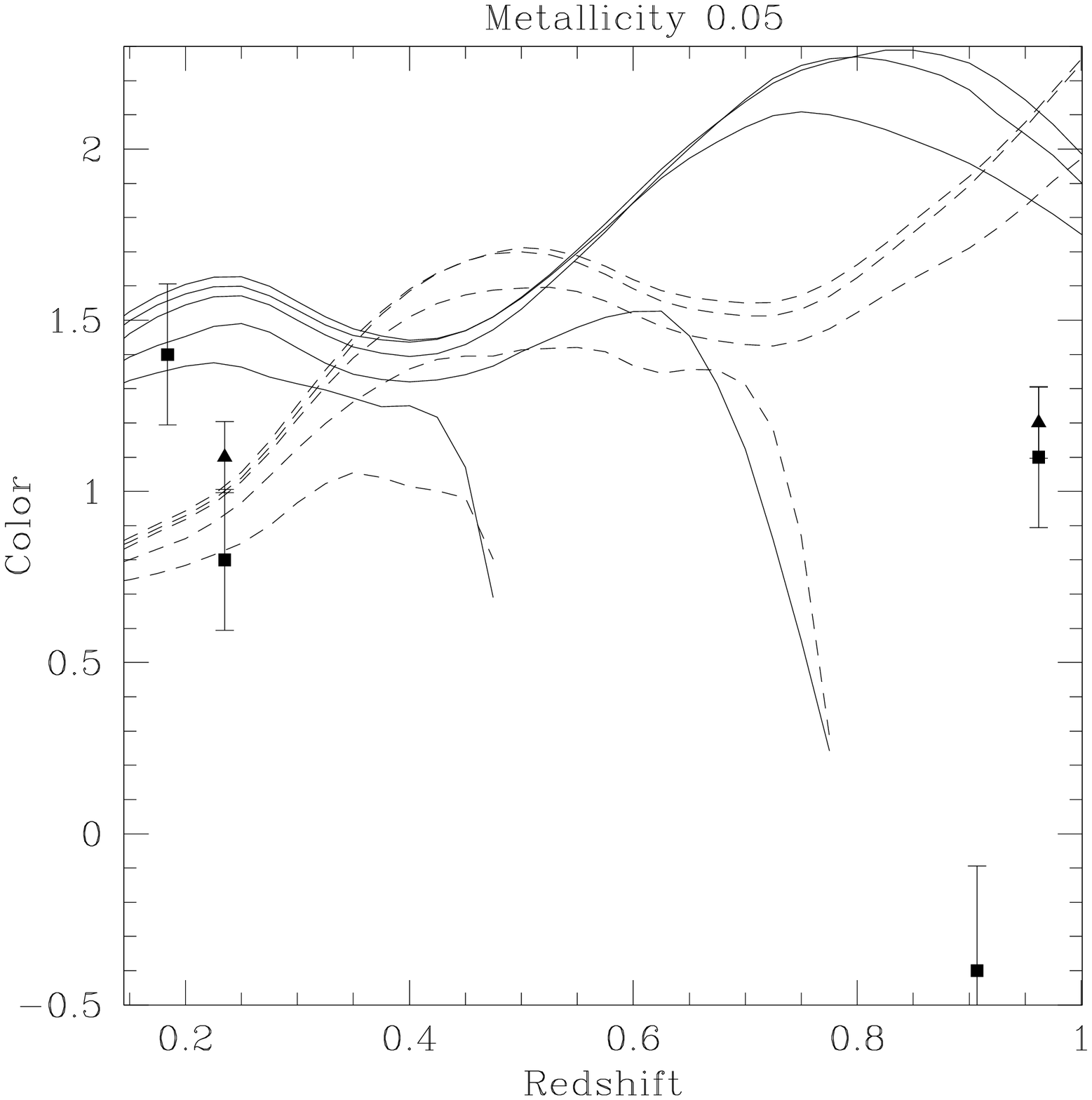}
\includegraphics[width=0.7\columnwidth]{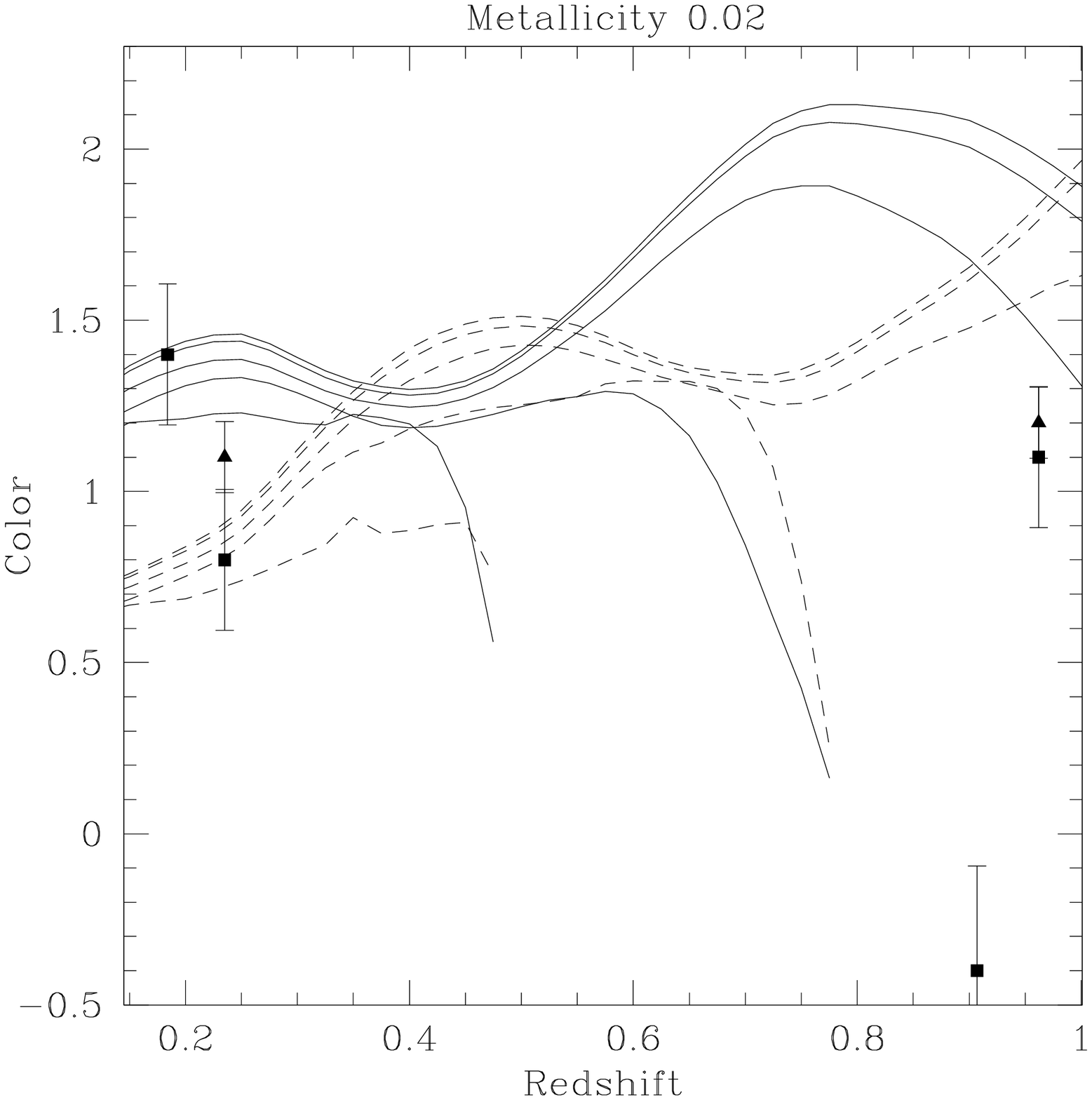}
\includegraphics[width=0.7\columnwidth]{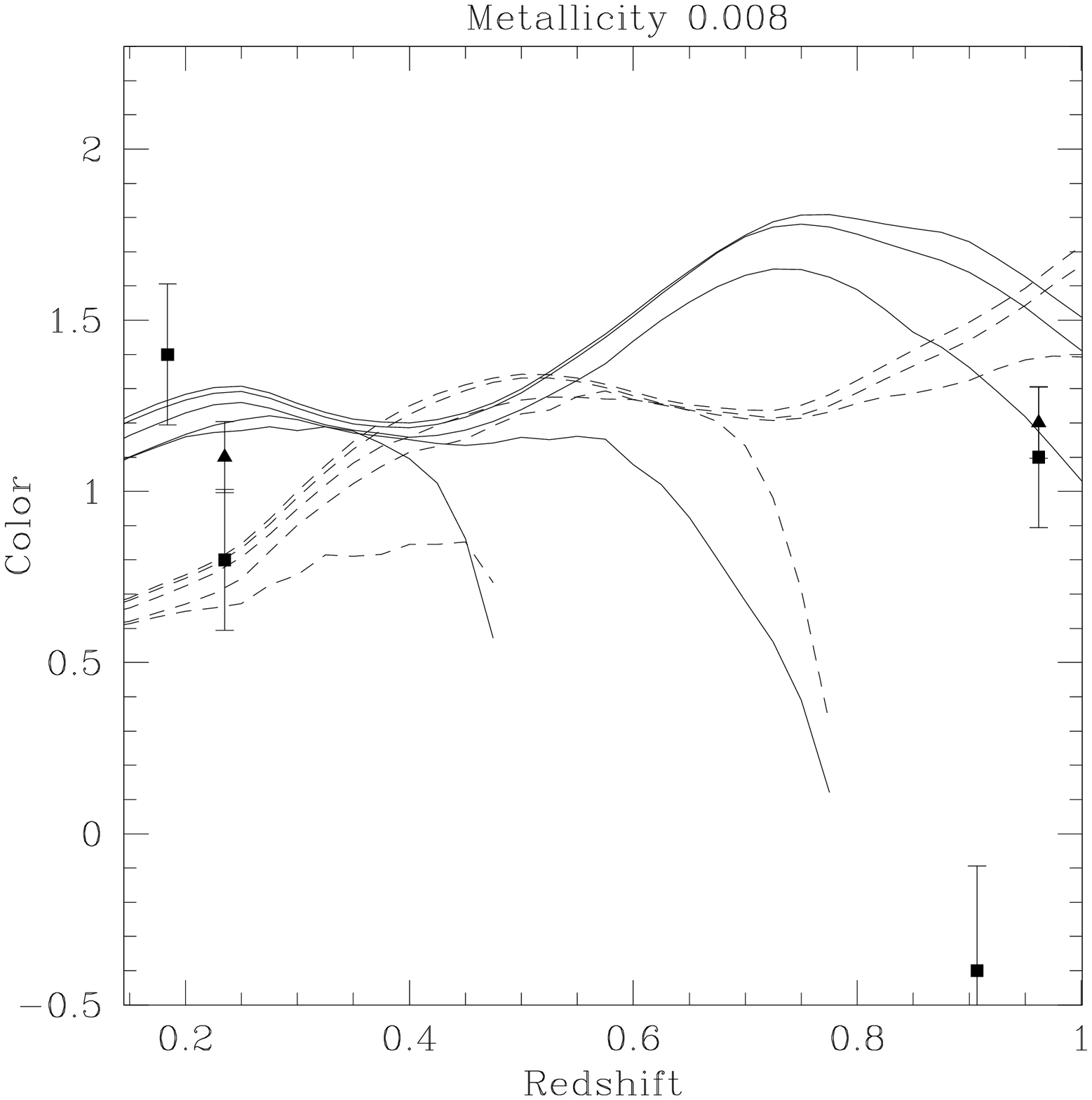}
\caption{Instantaneous, single burst stellar population synthesis models (lines) and observational colors (points) for our data. Squares and solid lines represent $B-V$, triangles and dashed lines represent $V-R$. There are five ages represented for each color: 5.0, 6.8, 9.3, 11.5 and 12.5 Gyr. The age of the population  increases from bottom to top. The discrepant data point corresponds to \object{B0316+161} (at {\it z} = 0.907). However, the measured color of \object{B0316+161} may be affected by bright nearby objects.}
%PONER TODAS LAS EDADES DONDE DICE QUE CRECEN Y YA VALE? Los redshifts son=0.5 , 0.8, 1.5 , 3.0, 5.0, las edades en ese orden: 5, 6.8,9.3,11.5 12.5 Gyr. Especificar que es SSP}
\label{sspm}
\end{figure}

The optical color data are in better agreement with the passively evolving instantaneous burst models 
(Figure \ref{sspm}), consistent with previous works \citep[e.g.,][]{Vries00}. 
Our data seem to favor models with ages of $\sim$12 Gyr and metallicity 0.008. However, the 
known degeneracy between metallicities and ages suggests that these results should be viewed with
caution. In addition, we note that our HST UV observations suggest that many GPS hosts
contain a small contribution from recent star formation \citep{Labiano05ACS, Labiano07ACS}. %(Labiano et al 2006). 

%The probable contribution from emission line gas and this

\section{Notes on individual sources}
\label{3sec:notes}

\begin{figure*}[p]
\centering
\includegraphics[width=\columnwidth]{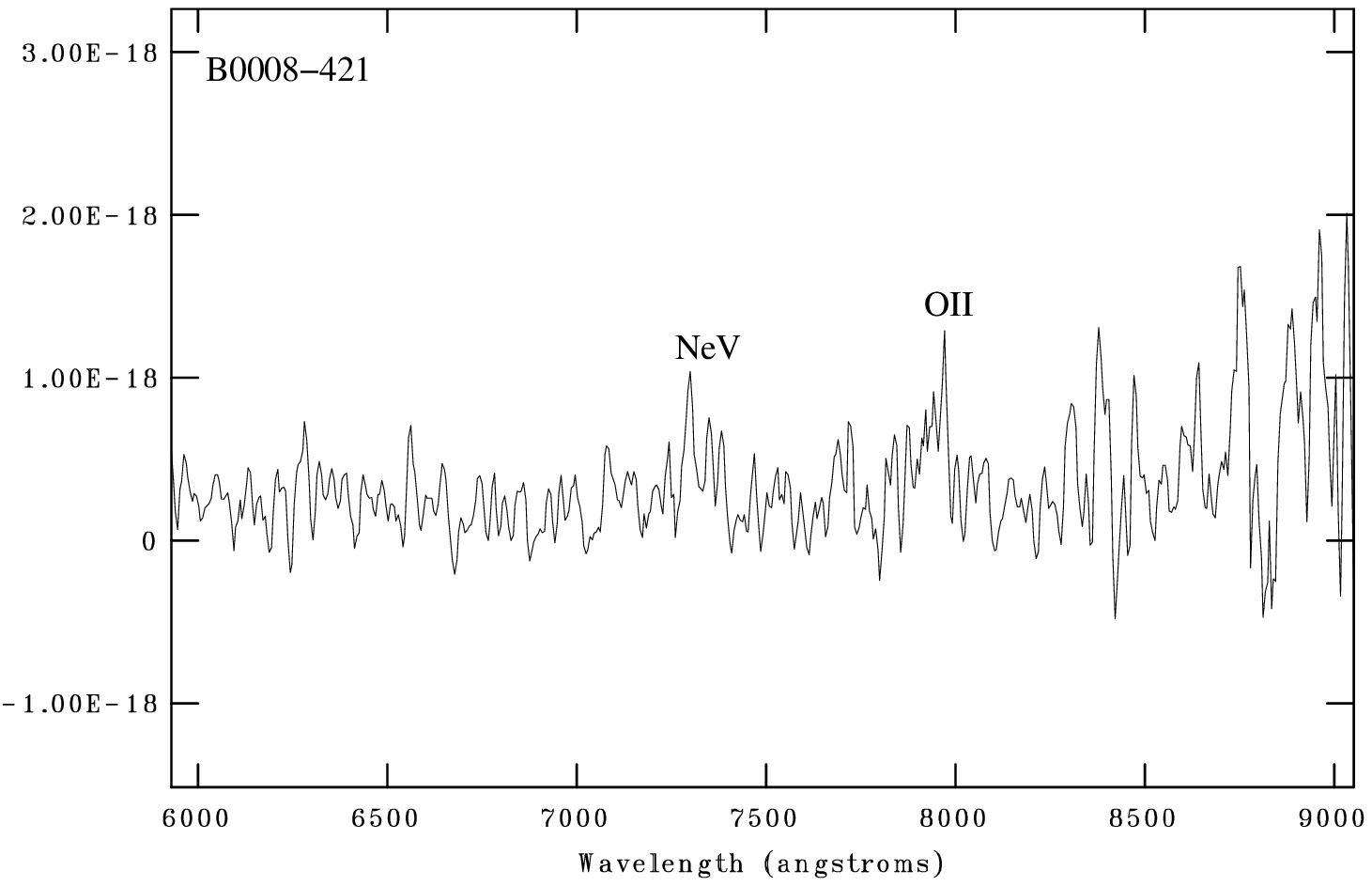}  \hfil \includegraphics[width=\columnwidth]{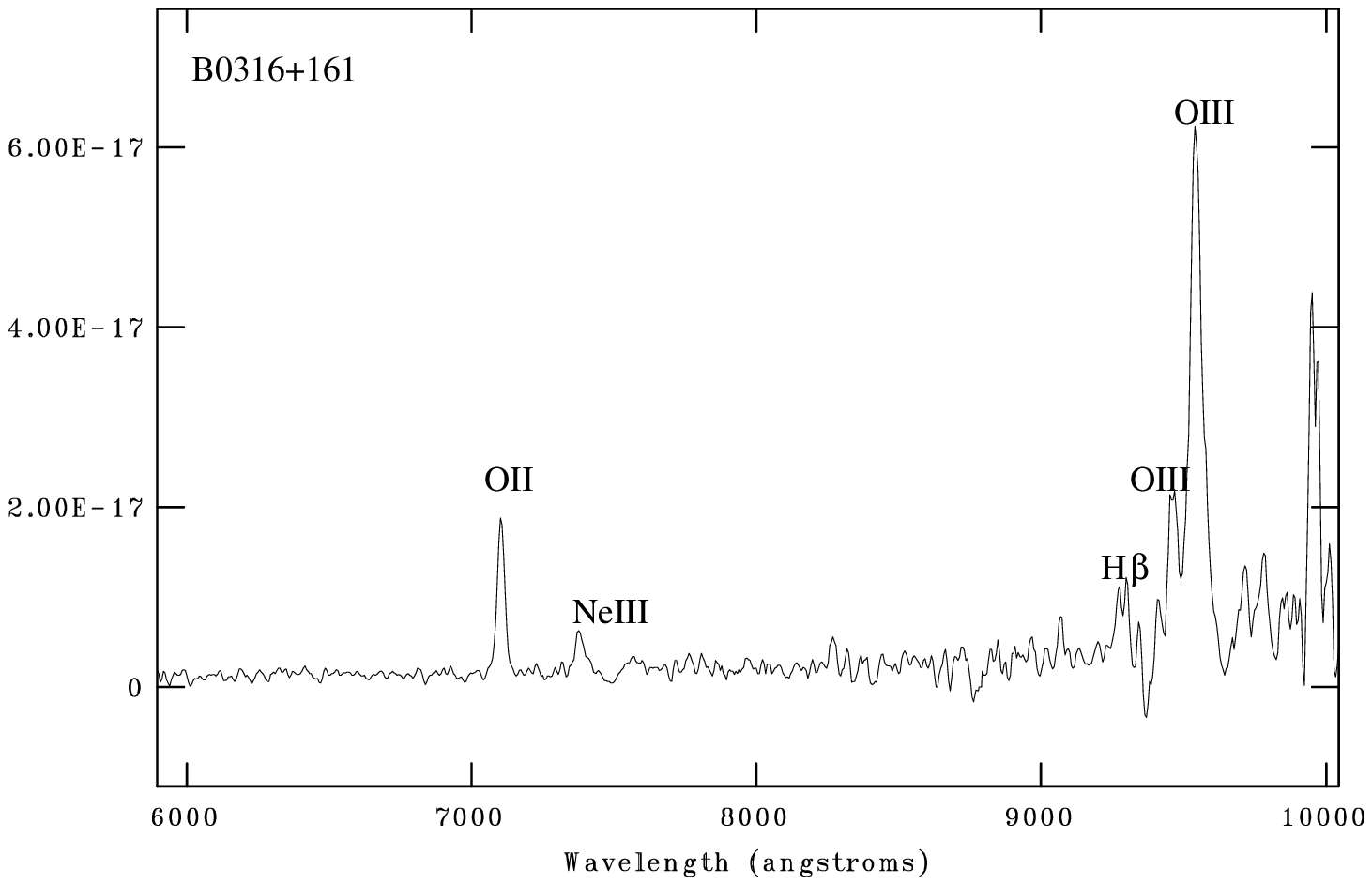}
\includegraphics[width=\columnwidth]{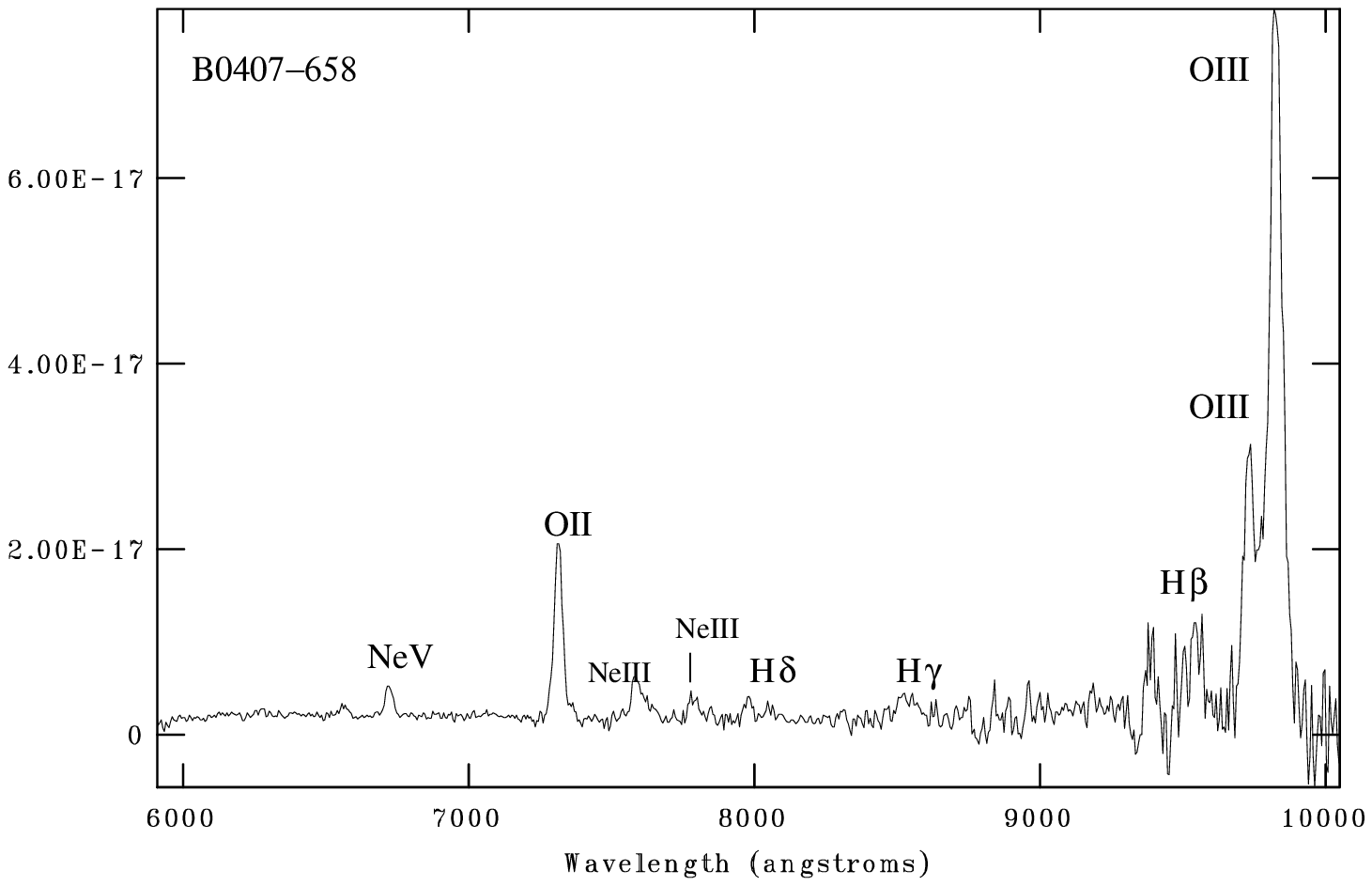} \hfil \includegraphics[width=\columnwidth]{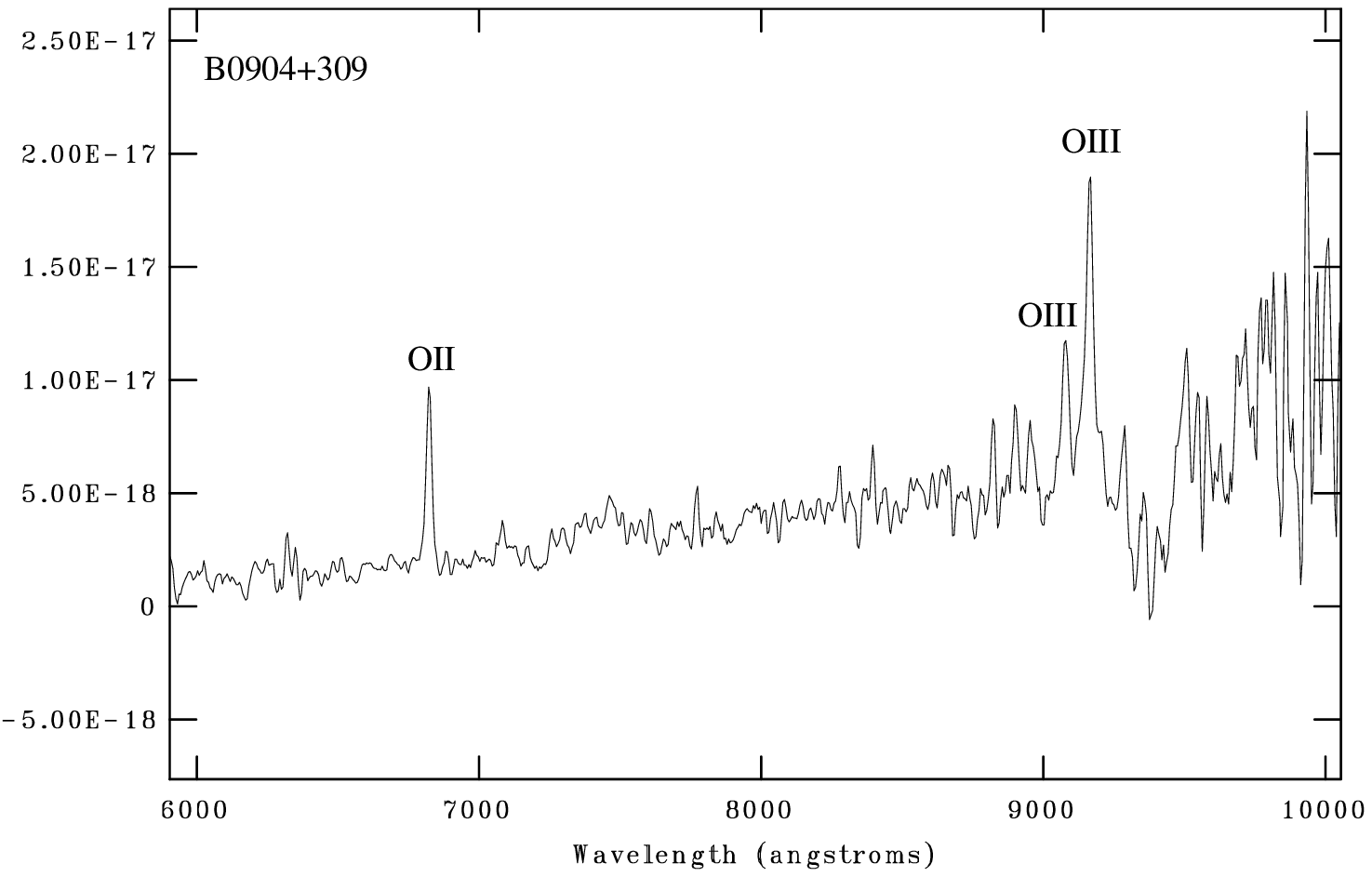} 
\includegraphics[width=\columnwidth]{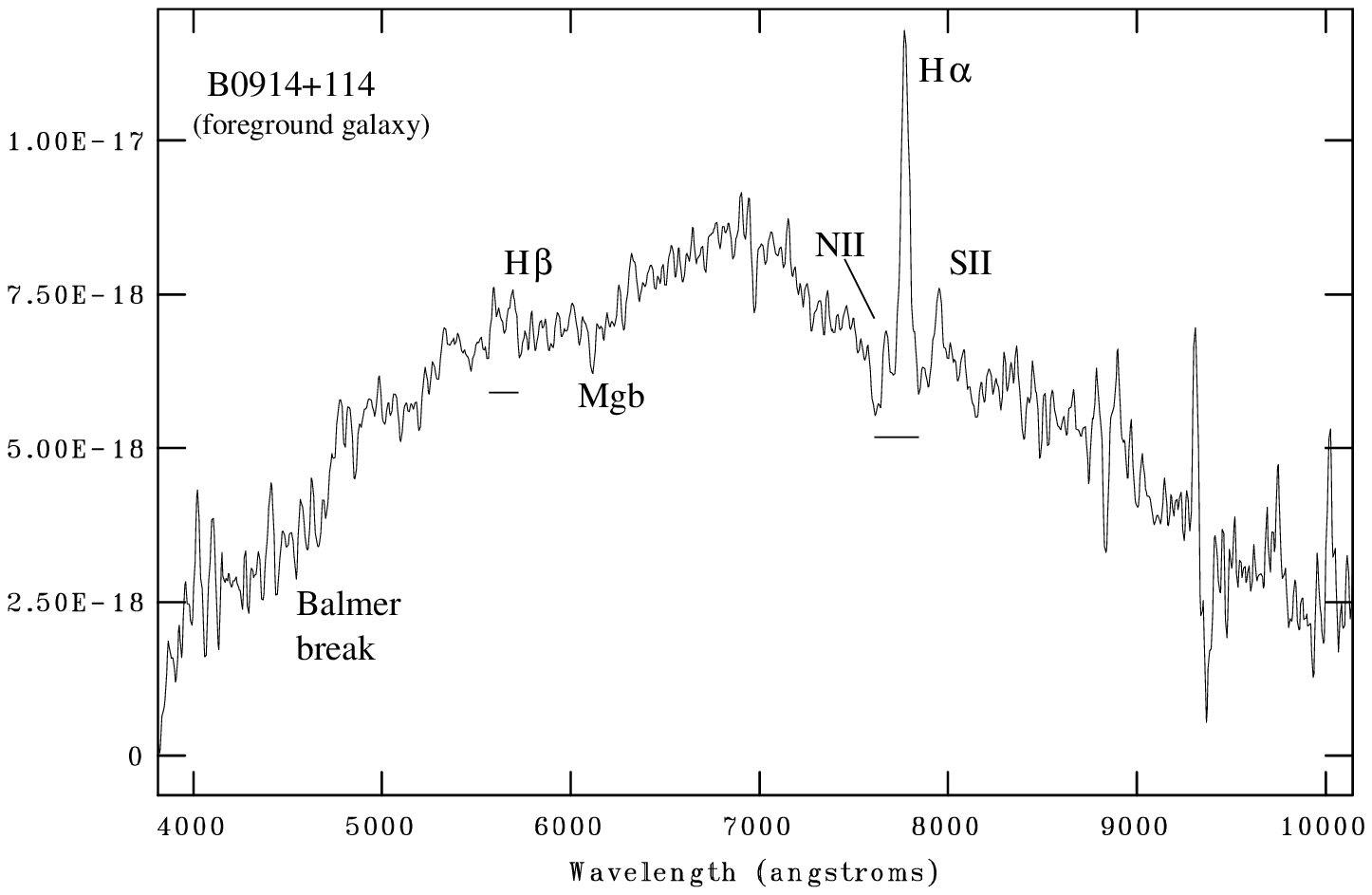} \hfil \includegraphics[width=\columnwidth]{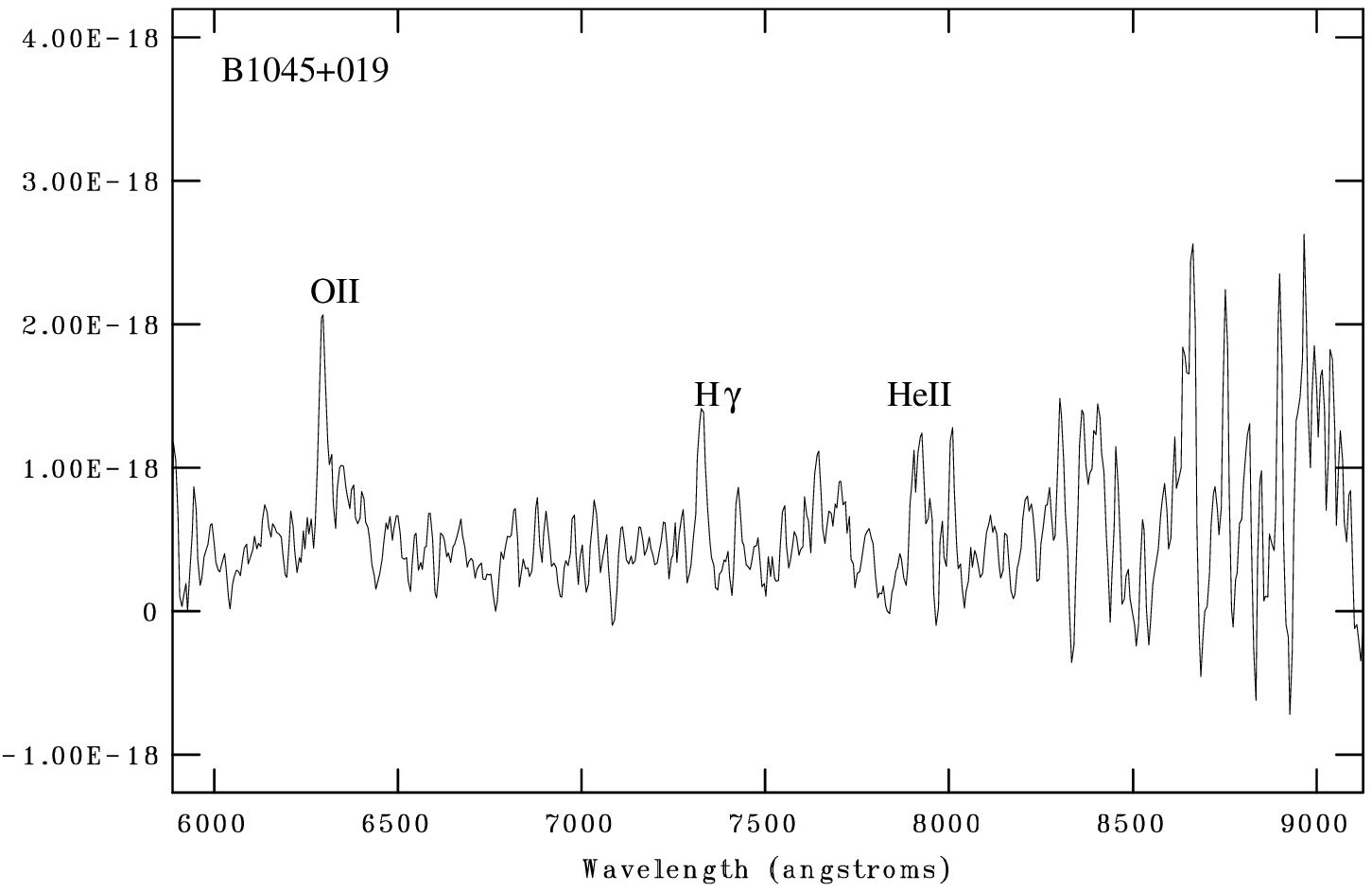}
\includegraphics[width=\columnwidth]{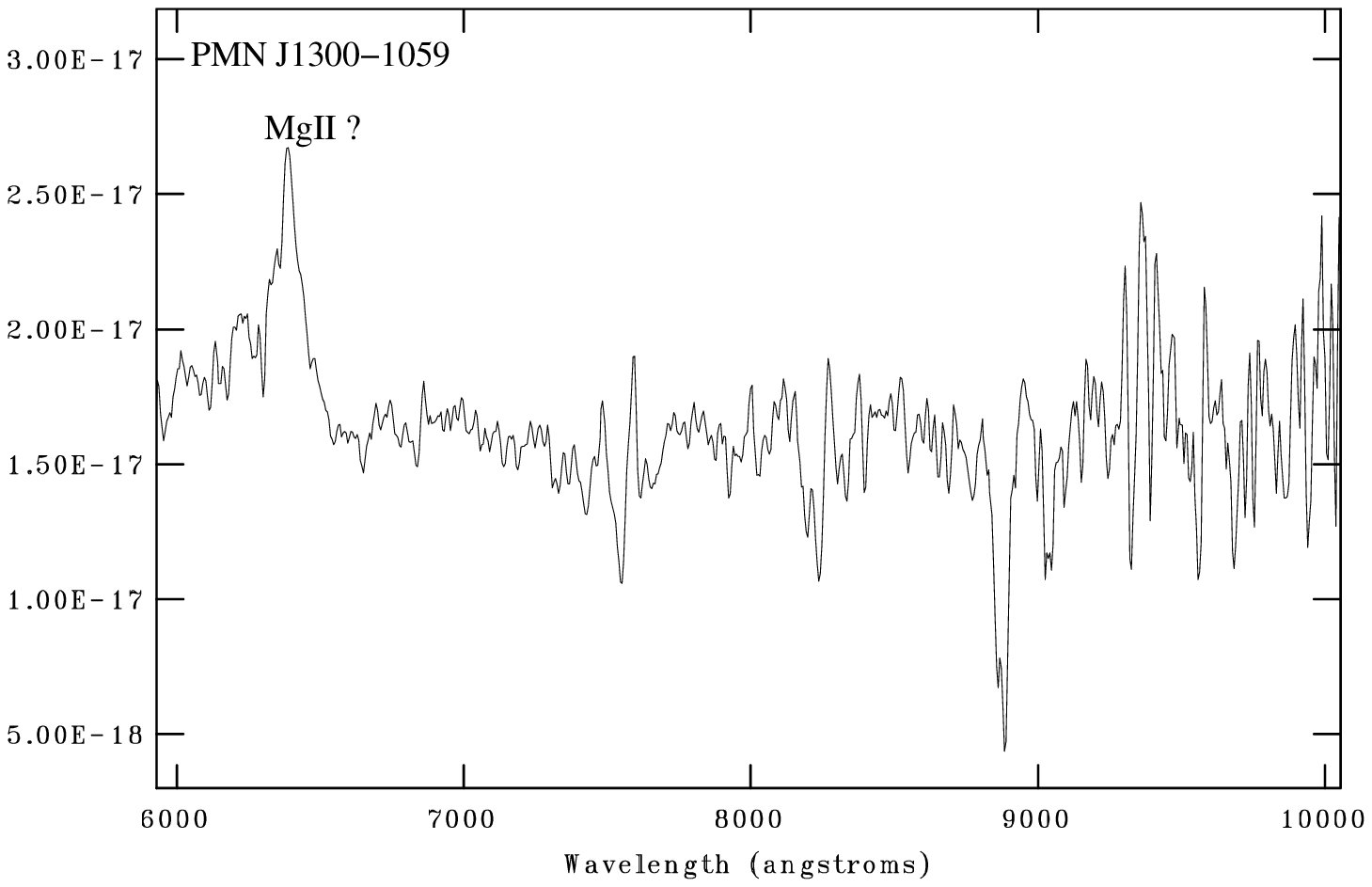} \hfil \includegraphics[width=\columnwidth]{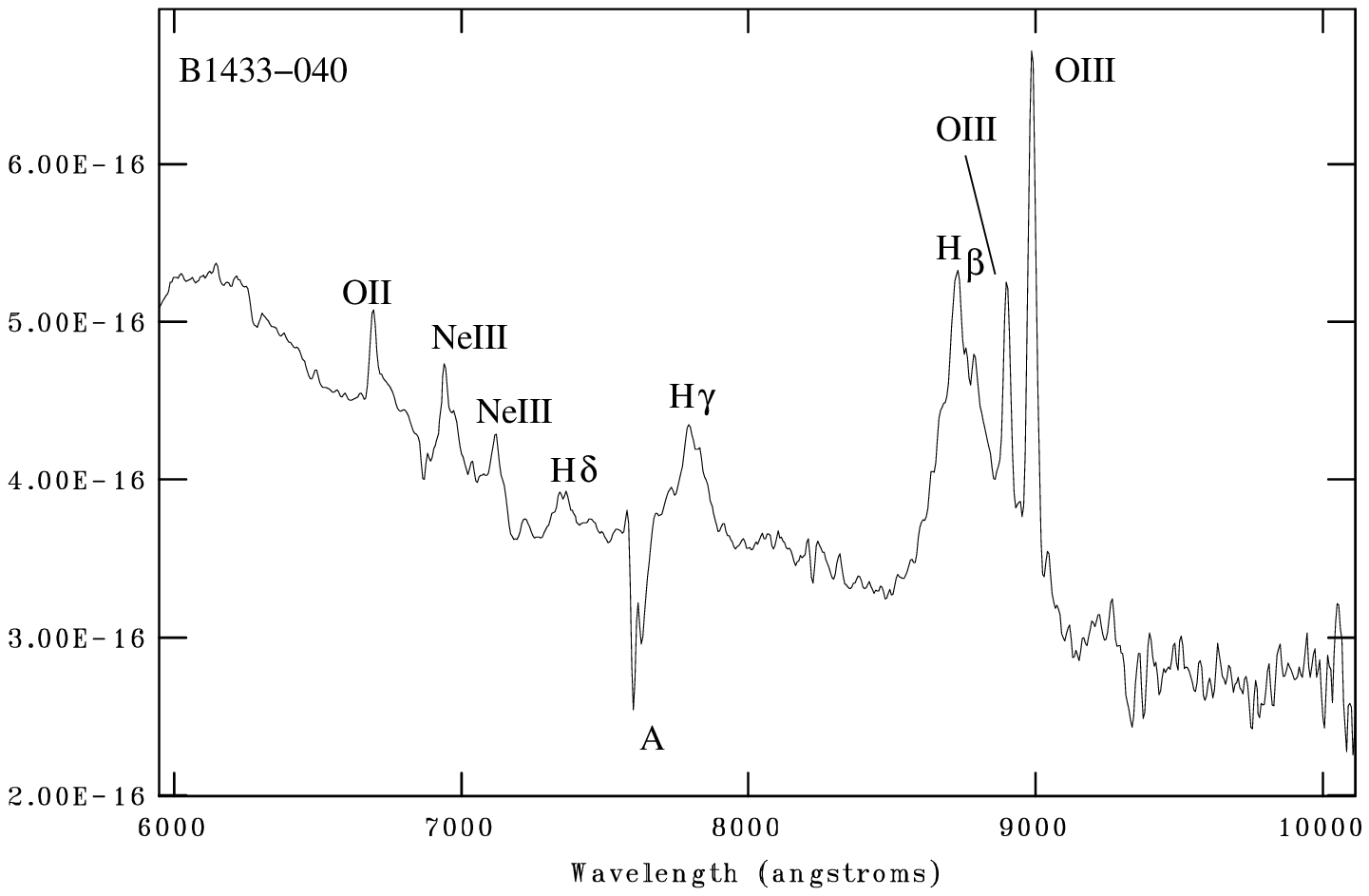}
\caption{Spectra of the sources with identified emission lines. Flux is in erg cm$^{-2}$ s$^{-1}$ ~\AA$^{-1}$. \object{B0914+114}: The spectrum corresponds to the previous -incorrect- identification (see text for details). The horizontal lines show an estimation of the possible FWHM of the absorption Hydrogen lines. }
\label{spec}
\end{figure*}

\begin{figure*}[p]
\centering
\includegraphics[width=\columnwidth]{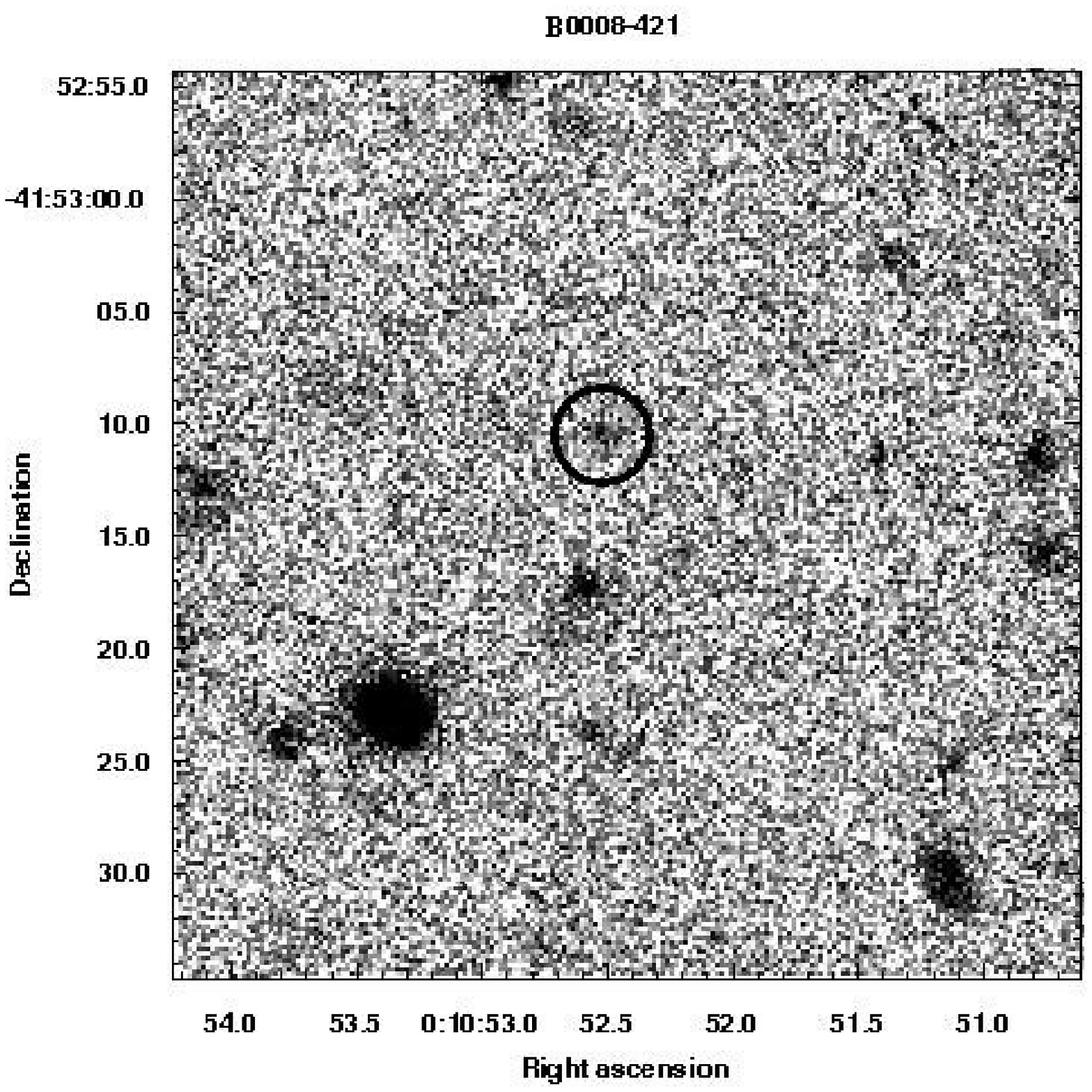} \hfill \includegraphics[width=\columnwidth]{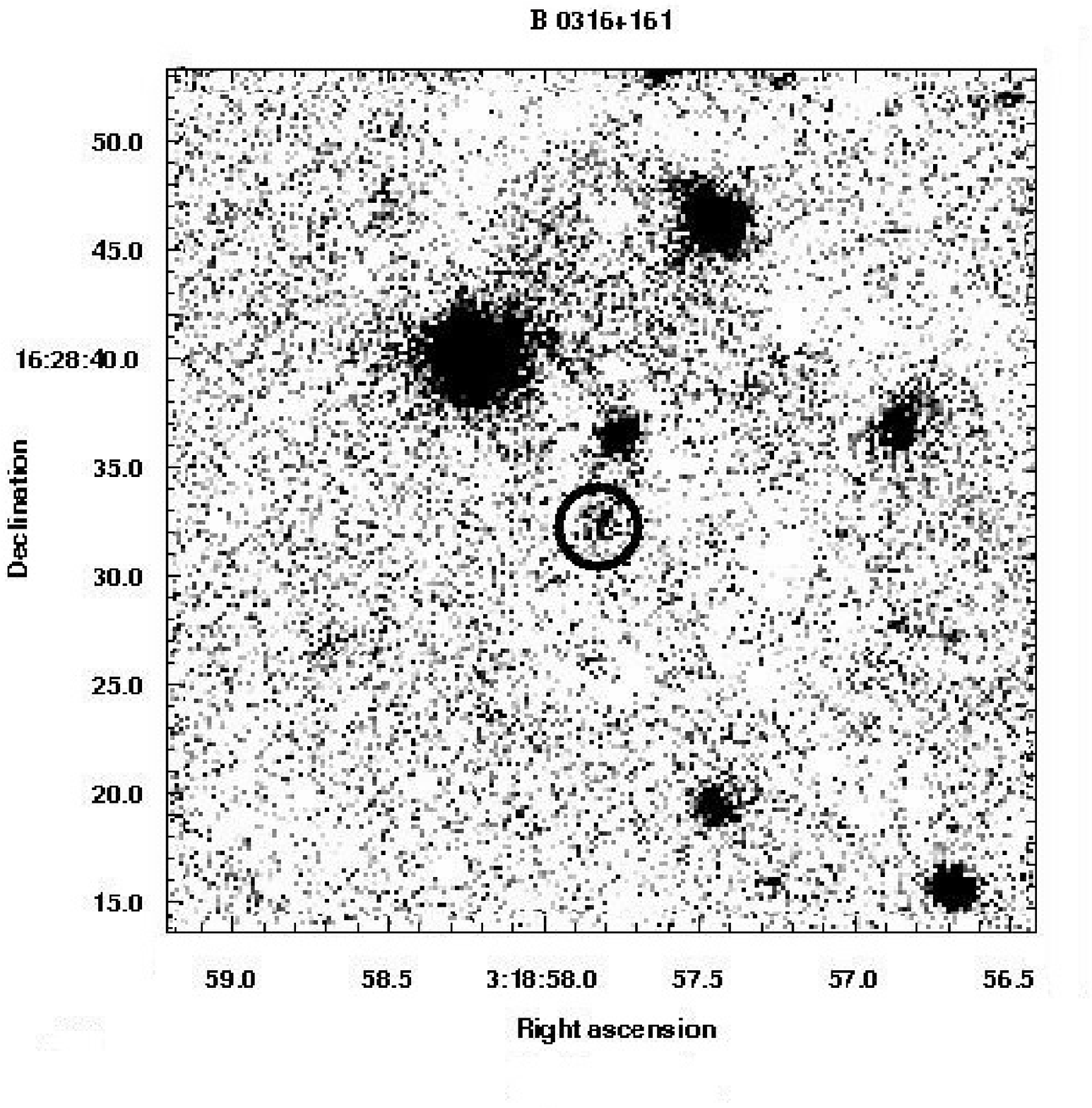}
\includegraphics[width=\columnwidth]{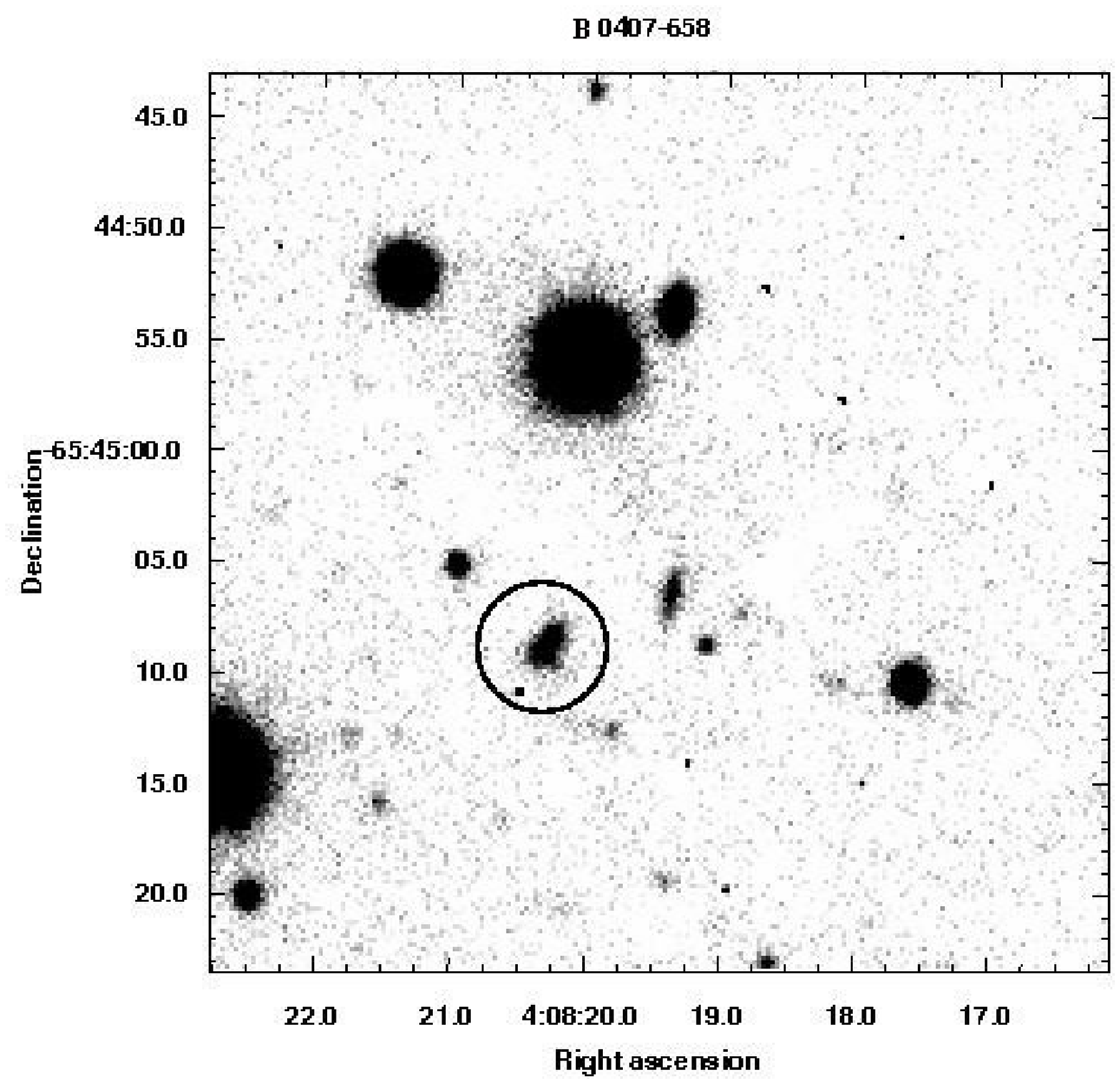} \hfil %\includegraphics[width=0.7\columnwidth]{0554-026.map.eps}
\includegraphics[width=\columnwidth]{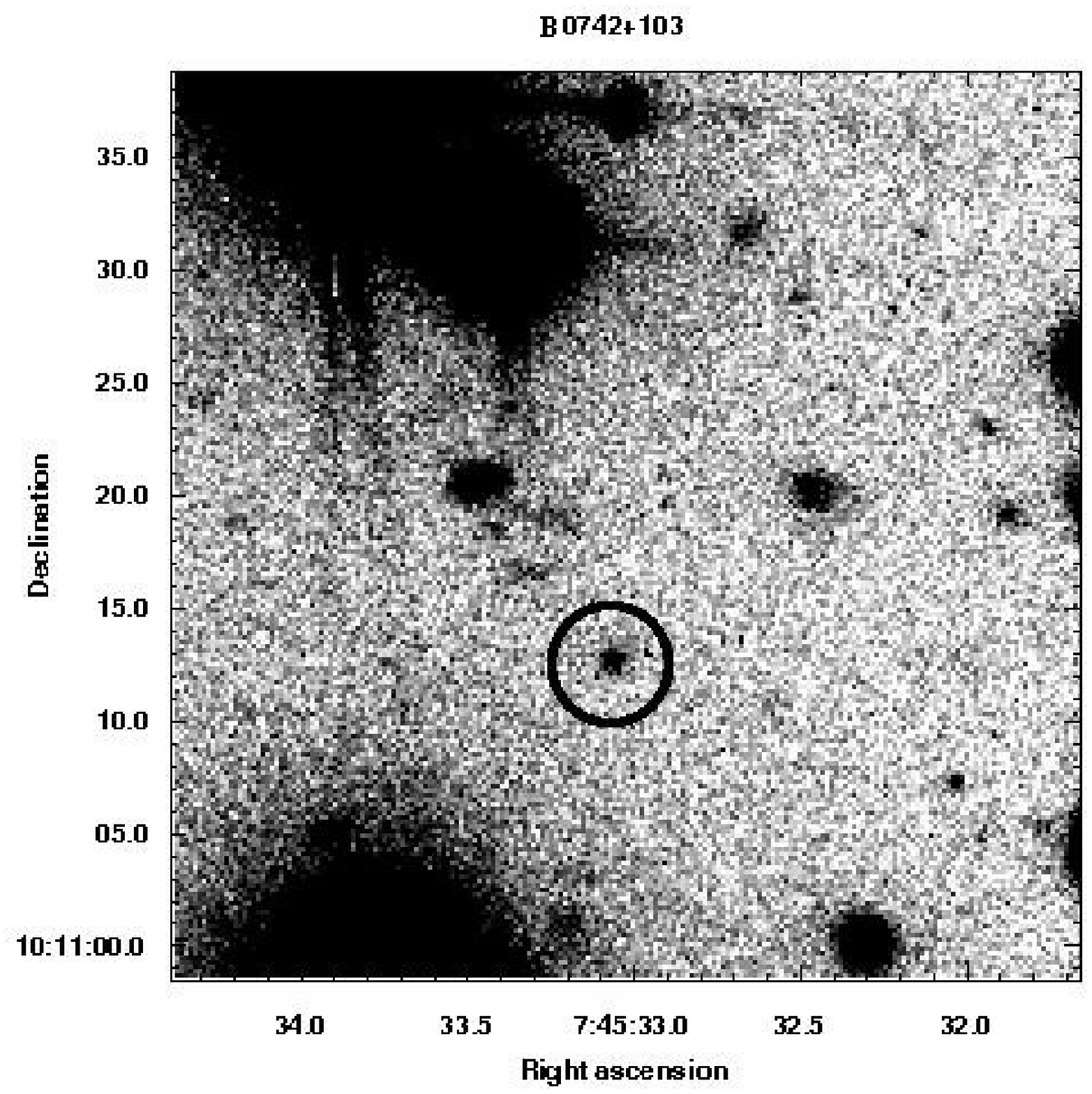}
\includegraphics[width=\columnwidth]{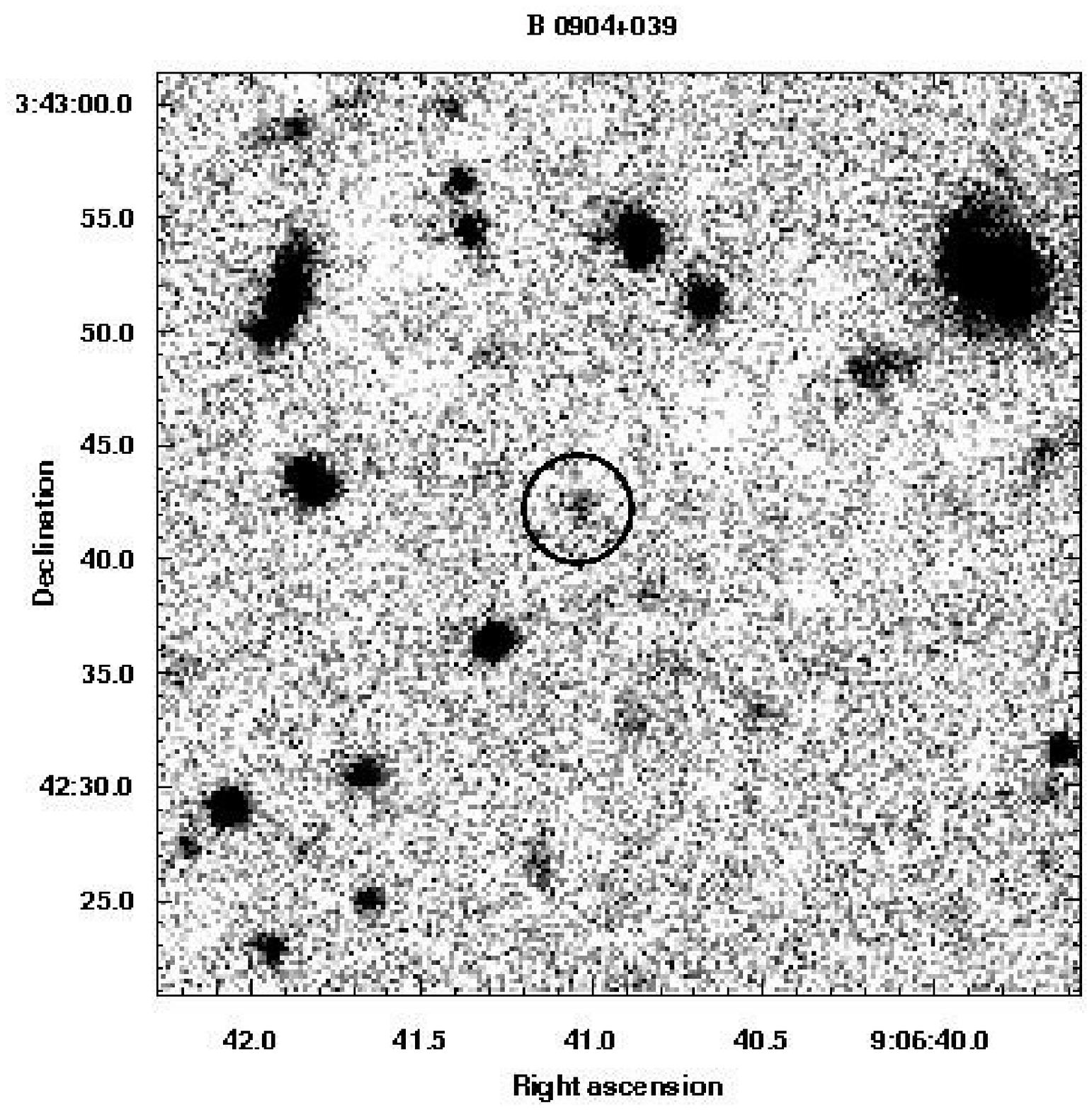} \hfil \includegraphics[width=\columnwidth]{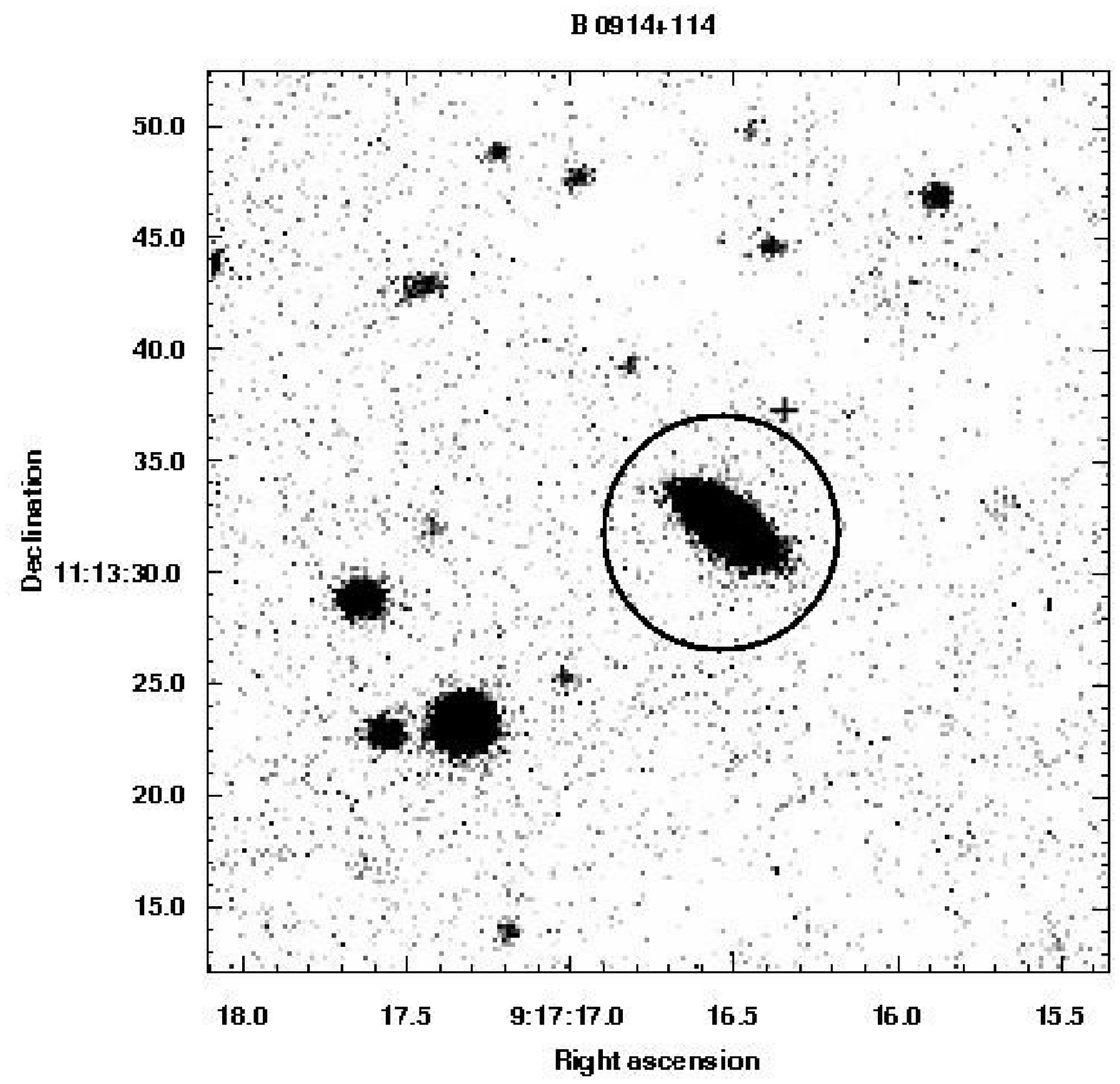}

\caption{Finding charts for the identified sources. The images correspond to the band of our VLT observations were sources were brighter. The radio position of 0914+114 is 6\arcsec\ north of the observed galaxy (circled) which was previously - and incorrectly  - identified as the optical counterpart (see text). The radio position is marked with a cross. }
\label{maps}
\end{figure*}

%%%% January.
{\bf \object{B0008--421}}: Following unsuccessful attempts by \citet{Serego94}, \citet{Vries95} and \citet{Costa01}, we have now identified the bright radio source \object{B0008--421} with an $R$=24.3, somewhat diffuse galaxy ({Figure \ref{maps}}). This magnitude is consistent with previous non-detections (which had detection limits down to magnitude 23). We detect a faint object, 0.2\arcsec\ from the VLBI position. The spectrum is consequently faint and very noisy making it difficult to distinguish emission lines from noise ({Figure \ref{spec}}). Comparison with the GPS Hubble diagram \citep[][ see Figure \ref{Hubble}]{O'Dea96} shows that for an apparent $R$ magnitude $\sim$24, the expected redshift is between 1 and 2. We find two dubious emission lines consistent with this range of redshift: [\ion{Ne}{v}]$\lambda$3425 and [\ion{O}{ii}]$\lambda$3727 at 7294 \AA\ and 7940 \AA, which would correspond to a redshift of {\it z} = 1.130. 
%a redshift between 0.4 and 0.5, we would expect an apparent $R$ magnitude around 19 or 20. According to the diagram, a magnitude $\sim$ 24 corresponds to a GPS galaxy at redshift between 1 and 2.

{\bf \object{B0316+161}}: Our deep image ({Figure \ref{maps}}) confirms the earlier identification of this well known GPS radio source (also known as CTA21), by \citet{Stanghellini93}. We find a rather compact host, and note that the object seen 8\arcsec\ NNW of the \object{CTA21} identification in the Stanghellini et al. image is unrelated. The relative faintness of \object{B0316+161}, and the proximity of bright objects, may be affecting our measured magnitudes. The spectrum shows a weak continuum with bright [\ion{O}{ii}] and [\ion{O}{iii}] lines ({Figure \ref{maps}}). We measure {\it z}= 0.907 based on five lines. 

{\bf \object{B0407--658}}: \citet{Stickel96} identified the optical counterpart of this radio source 
as a galaxy. The spectrum ({Figure \ref{spec}}) shows a faint continuum spectrum where we identify nine emission lines 
at {\it z} = 0.962. The emission line gas is more extended (3.2$\arcsec$, 25 kpc) along the slit than 
the stellar continuum. We observe no shift in wavelength along the spatial direction in the 
spectra. Our resolution ($\sim18$ \AA) yields a velocity resolution limit of $\sim 500$ km/s 
for the central part of the spectrum. Recent radio spectral data (NED)\footnote{NASA/IPAC Extragalactic 
Database.} show no peak around 1 GHz for this source, suggesting a CSS or larger radio source. 
However, the source is unresolved in ATCA observations \citep[resolution $\sim$5x3\arcsec\,][]{Morganti93}.

%With this redshift, the emission line gas extends extends 3.2\arcsec\ = 25 kpc along the slit. 
 
%\input{0408spec.tex}

{\bf \object{B0437--454}}: The spectrum shows a bright continuum but no emission or absorption lines. 
Updated radio data indicate that the radio spectrum of \object{B0437--454} is only marginally peaked. 
In addition, pronounced variability was reported: 0.6 Jy vs. 1.4 Jy at 5 GHz 
\citep[e.g.,][]{Wright94, Wright90}. Also, given the identification with an optical point source, 
we conclude that the object is {probably} a BL Lac object and should be removed from the GPS list. 
%One possible line at 9539 \AA, probably a cosmic ray. If it were [\ion{O}{iii}]  (at  z= 0.905), we would not be detecting [\ion{O}{ii}] 3727 (which should be at 7100 \AA \, for that redshift). 
% Quito la linea anterior por Isa.

{\bf \object{B0554--026}}: We confirm the identification of \citet{Vries00b} ({\it z}=0.283) of this galaxy. We find a rather bright ($B$=18.3, $V$=17.5) and extended source ($\sim$5\arcsec). We obtain $R$ = 16.4 but this band may be affected by calibration problems (see Section \ref{subsec:mags}).

{\bf \object{B0742+103}}: Neither \citet{Stickel96} nor \citet{Vries95} were able to identify the host of 
this relatively bright radio source. Our UT1 image ({Figure \ref{maps}}), which unambiguously identifies the source with a 
compact host galaxy, indicates that the tentative identifications by \citet{Fugmann88} and  
\citet{Vries00b} were correct: we establish $R$=23.1. We obtained long slit spectra of this source both in the January and 
the December run. Both observations show a faint spectrum and features which could be lines. 
However, none of these features are present in both spectra. \citet{Best03}  measure 2.624 $\pm$0.003 based on \ion{Ly}{$\alpha$}, \ion{C}{iv}, \ion{He}{ii} and \ion{C}{iii}] between 4400 and 6920 \AA. We detect a faint (slightly brighter than noise) emission at 6922 \AA ~ which could be the \citet{Best03} \ion{C}{iii}]. Having only one emission feature, we cannot get an independent redshift measurement. If their redshift is correct, using the 5 GHz flux density from \citet{O'Dea98}, the source would have a radio power of 2.1$\times10^{28}$ W/Hz. %Best03 : Resol es 12A y toman dos espectros de 1200s cada uno

{\bf \object{B0904+039}}: The deep UT1 image ({Figure \ref{maps}}) reinforces the earlier identification of this GPS radio source \cite[][ $I$=22.5]{Vries00b} with a faint host ($V$=24.9) in a group of faint galaxies. The spectrum shows a weak continuum ({Figure \ref{spec}}). We measure {\it z}=0.830 based on the [\ion{O}{ii}] and [\ion{O}{iii}] lines.

{\bf \object{B0914+114}}:  Our $B$ and $V$ band images show an empty field at the accurate
FIRST survey radio coordinates ({Figure \ref{maps}}). Based on uncertain WSRT coordinates, \citet{Stanghellini93} suggested that 
the disk galaxy $\sim$6\arcsec\ south of the current radio position could be the host of the radio source. 
Thus, this disk galaxy is not the optical counterpart of the GPS source. We obtained a spectrum of the 
disk galaxy, which shows a faint stellar dominated continuum ({Figure \ref{spec}}). We observe narrow H$\alpha$ and 
H$\beta$ in emission on top of a broad absorption which also may be affecting [\ion{N}{ii}] and 
[\ion{S}{ii}] emission. Averaging the redshifts of all the observed features, we obtain {\it z}=0.178. 
The H$\alpha$ emission at 7772~\AA ~had been detected before \citep{Vries00b}.

{\bf \object{B1045+019}} The weak continuum and noisy spectrum makes it difficult to distinguish real emission lines from noise. We only find one --dubious-- possibility for redshift ({\it z}=0.689). If this redshift is correct, we are not detecting H$\beta$ and [\ion{O}{iii}] 5007 (at 8200 and 8550~\AA ~ respectively). Radio observations were discussed in \citet{Vries00b} suggesting that this radio source may not be a GPS.

{\bf \object{PMN J1300--1059}}: This source is from a list of candidate GPS sources found by comparing 
VLA NVSS and WSRT WISH survey data. We find one emission line at 6385~\AA. The emission may consist of a narrow ($\sim80$ \AA) and a broad ($\sim200$ \AA) component but the edge of the chip is too close to deblend it accurately.  This line could be \ion{Mg}{ii} $\lambda$ 2799 at {\it z}=1.283, and in that case the [\ion{O}{ii}] doublet at 3727~\AA \, is not detected. If it were \ion{C}{iv}$\lambda 1549$, we would expect \ion{C}{iii}]$\lambda 1909$ at 7826~\AA. The relatively bright continuum suggests a QSO, which would be consistent with broad \ion{Mg}{ii} .

% \input{WhittleMod}
%\begin{figure}
%\centering
%\label{WhittleMod}
%\end{figure}

{\bf \object{B1433--040}}:  \citet{Vries00b} already drew attention to the fact that the GPS source \object{B1433--040} should not be identified with the considerably brighter radio source \object{4C~--04.51}. The optical spectrum shows a very strong continuum with broad ($\sim$ 200~\AA) and narrow ($\sim$30~\AA) emission lines ({Figure \ref{spec}}). The spectral shape and presence of bright broad lines is consistent with a QSO. We observe a strong asymmetry in the broad emission. The asymmetry index \citep[AI20,][]{Heckman81} is defined as (WL20--WR20)/(WL20+WR20), where WL20 and WR20 are the half width of the line to the left (WL20) and right (WR20) at the 20\% intensity level. We measure AI20$\sim$0.35, towards the red,  for H$\beta$ (and H$\gamma$), which is large, but not unusual. This asymmetry can be explained by inflow or outflow of gas, together with a line opacity (or scatter) cloud which blocks the emission at one side of the AGN \citep[e.g.,][ and references therein]{Whittle85}. The optical spectrum of this radio source displays hydrogen emission lines of striking velocity width: we measure 28000 km/sec FWZI for H$\beta$, and note in addition its double-peaked nature.  The radio spectrum {has a relatively broad} peak around 1 GHz \citep[e.g.,][]{Spoelstra85}. However, more observations, especially at frequencies $\lesssim$1 GHz, are needed. There seems to be some variability in the 408 MHz and $\sim$1.4 GHz \citep{Large81,Wright90, White92} observed flux densities. Given the unresolved optical host, we suggest that the optical counterpart of  \object{B1433--040}  is a quasar at {\it z}=0.796. %Although the literature (NED) radio spectrum seems to peak around 1 GHz, the high flux observed at 178 MHz suggests that \object{B1433--040} is not a GPS source \citep[e.g.,][]{Douglas96,Wright90}. It may be a flat spectrum quasar where Doppler boosting is affecting the spectrum. PARECE QUE ALGUNOS PUNTOS CORRESPONDEN  A 4C --04.51.
 
%; see Figure \ref{WhittleMod}
%\input{1433spec.tex}

{\bf \object{B1601--222}}: Featureless, very noisy spectrum with a moderately bright continuum. \citet{Snellen02} measure {\it z} = 0.141 based on the absorption features: G-band 4300, H$\beta$ 4861, Mg b 5169 and Na D 5899, which correspond to $\sim$4900--6800~\AA. We covered the range between 5700 and 9200~\AA ~ and observe no features. However, their spectrum has higher wavelength resolution and longer exposure time (2700 seconds).
%(H$\alpha$ with their redshift would be at 7488~\AA). 

{\bf \object{B1648+015}}: Featureless with a not very bright continuum. \citet{Stickel96} identifies this source as a quasar. The radio spectrum shows variability so it is probably {either} a flat spectrum quasar or a BL Lac object.

\section{Master list}
We have updated the \citet{O'Dea91} GPS sources master list with new available data (Table \ref{masterlist}). This new list consists of those radio sources with intrinsic turnover frequency between 0.4 and 5 GHz (GPS) or higher (High Frequency Peakers, HFP). For sources with unknown redshift, we have extended the selection down to 0.3 GHz.
We expect this list to be useful for workers in the field of
radio galaxies and quasars in general, and compact radio sources and AGN 
hosts in
particular. The list is furthermore electronically available via URL http://www.damir.iem.csic.es/extragalactic/people/labiano/ and http://www.cis.rit.edu/$\sim$cposps/

% compiled a master list of GPS radio sources from the literature and updated the  
%Samples we use, las recopiladas y corregidas por:\\
%\cite{Augusto06, Snellen00, Vries00, O'Dea98, O'Dea91, Vries 95, Stanghellini98, Tziomis02, Fanti01, Tschager03b, Stanghellini01, Taylor03, Marecki99, Tinti05b, Kunert05, Snellen02, Edwards04} Y otros. Poner criterios seleccion 0.4--5GHz.
%Por supuesto, en caso de conflicto, usamos la mas actual
%METER TABLA!

\section{Summary}

We presented VLT deep optical imaging and spectroscopy targeting the host galaxies of GPS radio sources. The sample was comprised of unidentified objects from the master list of \citet{O'Dea91}, updated by \citet{Vries97b}.

We have found new optical counterparts (down to magnitudes $\sim$25) of GPS sources \object{B0008--421} and \object{B0742+103} and confirmed previous identifications of GPS sources: \object{B0316+161}, \object{B0407--658}, \object{B0554--026}, \object{B0904+039}. With new radio observations from the literature, we find that the radio spectra of \object{B0407--658}, \object{B0437--454} and \object{B1648+015} suggest that these sources are not GPS. However, high resolution radio observations are needed to confirm it. We do not detect the optical counterpart of 
\object{B0914+114} and suggest that previous identification corresponds to an unrelated galaxy 
(at {\it z}=0.178), 6\arcsec\ {South} of the current radio position.

We measure new redshifts for \object{B0316+161}, \object{B0407--658}, \object{B0904+039}, \object{B0914+114} (unrelated galaxy) and \object{B1433--040}, {and propose} uncertain redshifts for \object{B0008--421}, \object{B1045+019}, \object{PMN J1300--1059}. The following sources remain with undetermined redshift: \object{B0437--454}, \object{B0914+114} and \object{B1648+015}. We cannot confirm previous redshifts of: \object{B0742+103}, \object{B1601--222}. Our magnitudes seem to be consistent with previous measurements of GPS counterparts. 
%We find redder V--R colors for some sources but this can be due to problems in the calibration of the {\it R}-band observations. 

Stellar population synthesis models are inconsistent with constant or exponentially declining star formation in the host. The data generally agree with single instantaneous burst models in a passively evolving host, but do not yield useful information on age or metallicity.

%%\input{Spectra.tex}
%\input{Spectra1.tex} \clearpage
%\input{Spectra2.tex}
%%\input{FIGS/Maps.tex}

%%%%% FIG ?? %%%%%

%%%%%%%%%%%%%

\begin{acknowledgements}
A.L. wishes to thank Prof. dr. R.~F. Peletier (Kapteyn Astronomical Institute) for useful scientific discussions. This paper is based on observations made with ESO Telescopes at Paranal Observatory under Programmes 64.P--0482(A) and 68.B--0044(A)). This research has made use of NASA's Astrophysics Data System Bibliographic Services and of the NASA/IPAC Extragalactic Database (NED) which is operated by the Jet Propulsion Laboratory, California Institute of Technology, under contract with the National Aeronautics and Space Administration.
\end{acknowledgements}

\bibliographystyle{aa}
\bibliography{../ALORefs}

\begin{thebibliography}{66}
\expandafter\ifx\csname natexlab\endcsname\relax\def\natexlab#1{#1}\fi

\bibitem[{{Appenzeller} {et~al.}(1998){Appenzeller}, {Fricke}, {Furtig},
  {Gassler}, {Hafner}, {Harkl}, {Hess}, {Hummel}, {Jurgens}, {Kudritzki},
  {Mantel}, {Meisl}, {Muschielok}, {Nicklas}, {Rupprecht}, {Seifert}, {Stahl},
  {Szeifert}, \& {Tarantik}}]{Appenzeller98}
{Appenzeller}, I., {Fricke}, K., {Furtig}, W., {et~al.} 1998, The Messenger,
  94, 1

\bibitem[{{Augusto} {et~al.}(2006){Augusto}, {Gonzalez-Serrano},
  {Perez-Fournon}, \& {Wilkinson}}]{Augusto06}
{Augusto}, P., {Gonzalez-Serrano}, J.~I., {Perez-Fournon}, I., \& {Wilkinson},
  P.~N. 2006, \mnras, 368, 1411

\bibitem[{{Beasley} {et~al.}(2002){Beasley}, {Gordon}, {Peck}, {Petrov},
  {MacMillan}, {Fomalont}, \& {Ma}}]{Beasley02}
{Beasley}, A.~J., {Gordon}, D., {Peck}, A.~B., {et~al.} 2002, \apjs, 141, 13

\bibitem[{{Becker} {et~al.}(1991){Becker}, {White}, \& {Edwards}}]{Becker91}
{Becker}, R.~H., {White}, R.~L., \& {Edwards}, A.~L. 1991, \apjs, 75, 1

\bibitem[{{Begelman}(1996)}]{Begelman96}
{Begelman}, M.~C. 1996, in Cygnus A -- Study of a Radio Galaxy, 209

\bibitem[{{Best} {et~al.}(2003){Best}, {Peacock}, {Brookes}, {Dowsett}, {R{\"
  o}ttgering}, {Dunlop}, \& {Lehnert}}]{Best03}
{Best}, P.~N., {Peacock}, J.~A., {Brookes}, M.~H., {et~al.} 2003, \mnras, 346,
  1021

\bibitem[{{Bruzual} \& {Charlot}(2003)}]{Bruzual03}
{Bruzual}, G. \& {Charlot}, S. 2003, \mnras, 344, 1000

\bibitem[{{Chabrier}(2003)}]{Chabrier03}
{Chabrier}, G. 2003, \pasp, 115, 763

\bibitem[{{Condon} {et~al.}(1998){Condon}, {Cotton}, {Greisen}, {Yin},
  {Perley}, {Taylor}, \& {Broderick}}]{Condon98}
{Condon}, J.~J., {Cotton}, W.~D., {Greisen}, E.~W., {et~al.} 1998, \aj, 115,
  1693

\bibitem[{{Costa}(2001)}]{Costa01}
{Costa}, E. 2001, \aap, 367, 719

\bibitem[{{Dallacasa} {et~al.}(2000){Dallacasa}, {Stanghellini}, {Centonza}, \&
  {Fanti}}]{Dallacasa00}
{Dallacasa}, D., {Stanghellini}, C., {Centonza}, M., \& {Fanti}, R. 2000, \aap,
  363, 887

\bibitem[{{De Breuck} {et~al.}(2002){De Breuck}, {Tang}, {de Bruyn},
  {R{\"o}ttgering}, \& {van Breugel}}]{Breuck02}
{De Breuck}, C., {Tang}, Y., {de Bruyn}, A.~G., {R{\"o}ttgering}, H., \& {van
  Breugel}, W. 2002, \aap, 394, 59

\bibitem[{{de Ruiter} {et~al.}(1977){de Ruiter}, {Arp}, \& {Willis}}]{Ruiter77}
{de Ruiter}, H.~R., {Arp}, H.~C., \& {Willis}, A.~G. 1977, \aaps, 28, 211

\bibitem[{{de Vries} {et~al.}(1995){de Vries}, {Barthel}, \& {Hes}}]{Vries95}
{de Vries}, W.~H., {Barthel}, P.~D., \& {Hes}, R. 1995, \aaps, 114, 259

\bibitem[{{de Vries} {et~al.}(1997){de Vries}, {Barthel}, \&
  {O'Dea}}]{Vries97b}
{de Vries}, W.~H., {Barthel}, P.~D., \& {O'Dea}, C.~P. 1997, \aap, 321, 105

\bibitem[{{de Vries} {et~al.}(2000{\natexlab{a}}){de Vries}, {O'Dea},
  {Barthel}, {Fanti}, {Fanti}, \& {Lehnert}}]{Vries00}
{de Vries}, W.~H., {O'Dea}, C.~P., {Barthel}, P.~D., {et~al.}
  2000{\natexlab{a}}, \aj, 120, 2300

\bibitem[{{de Vries} {et~al.}(2000{\natexlab{b}}){de Vries}, {O'Dea},
  {Barthel}, \& {Thompson}}]{Vries00b}
{de Vries}, W.~H., {O'Dea}, C.~P., {Barthel}, P.~D., \& {Thompson}, D.~J.
  2000{\natexlab{b}}, \aaps, 143, 181

\bibitem[{{de Vries} {et~al.}(1998){de Vries}, {O'Dea}, {Baum}, {Perlman},
  {Lehnert}, \& {Barthel}}]{Vries98b}
{de Vries}, W.~H., {O'Dea}, C.~P., {Baum}, S.~A., {et~al.} 1998, \apj, 503, 156

\bibitem[{{De Young}(1997)}]{Young97}
{De Young}, D.~S. 1997, \apjl, 490, L55

\bibitem[{{di Serego-Alighieri} {et~al.}(1994){di Serego-Alighieri},
  {Danziger}, {Morganti}, \& {Tadhunter}}]{Serego94}
{di Serego-Alighieri}, S., {Danziger}, I.~J., {Morganti}, R., \& {Tadhunter},
  C.~N. 1994, \mnras, 269, 998

\bibitem[{{Edwards} \& {Tingay}(2004)}]{Edwards04}
{Edwards}, P.~G. \& {Tingay}, S.~J. 2004, \aap, 424, 91

\bibitem[{{Fanti} {et~al.}(1995){Fanti}, {Fanti}, {Dallacasa}, {Schilizzi},
  {Spencer}, \& {Stanghellini}}]{Fanti95}
{Fanti}, C., {Fanti}, R., {Dallacasa}, D., {et~al.} 1995, \aap, 302, 317

\bibitem[{{Fugmann} {et~al.}(1988){Fugmann}, {Meisenheimer}, \&
  {Roeser}}]{Fugmann88}
{Fugmann}, W., {Meisenheimer}, K., \& {Roeser}, H.-J. 1988, \aaps, 75, 173

\bibitem[{{Griffith} {et~al.}(1994){Griffith}, {Wright}, {Burke}, \&
  {Ekers}}]{Griffith94}
{Griffith}, M.~R., {Wright}, A.~E., {Burke}, B.~F., \& {Ekers}, R.~D. 1994,
  \apjs, 90, 179

\bibitem[{{Heckman} {et~al.}(1981){Heckman}, {Miley}, {van Breugel}, \&
  {Butcher}}]{Heckman81}
{Heckman}, T.~M., {Miley}, G.~K., {van Breugel}, W.~J.~M., \& {Butcher}, H.~R.
  1981, \apj, 247, 403

\bibitem[{{Kaiser} \& {Alexander}(1997)}]{Kaiser97b}
{Kaiser}, C.~R. \& {Alexander}, P. 1997, \mnras, 286, 215

\bibitem[{{Kaiser} {et~al.}(1997){Kaiser}, {Dennett-Thorpe}, \&
  {Alexander}}]{Kaiser97a}
{Kaiser}, C.~R., {Dennett-Thorpe}, J., \& {Alexander}, P. 1997, \mnras, 292,
  723

\bibitem[{{Kent}(1985)}]{Kent85}
{Kent}, S.~M. 1985, \pasp, 97, 165

\bibitem[{{Kuehr} {et~al.}(1981){Kuehr}, {Witzel}, {Pauliny-Toth}, \&
  {Nauber}}]{Kuehr81}
{Kuehr}, H., {Witzel}, A., {Pauliny-Toth}, I.~I.~K., \& {Nauber}, U. 1981,
  \aaps, 45, 367

\bibitem[{{Labiano} {et~al.}(2006){Labiano}, {O'Dea}, {Barthel}, {de Vries}, \&
  {Baum}}]{Labiano05ACS}
{Labiano}, A., {O'Dea}, C.~P., {Barthel}, P.~D., {de Vries}, W.~H., \& {Baum},
  S.~A. 2006, \nar, 50, 776

\bibitem[{{Labiano} {et~al.}(2007 in prep.){Labiano}, {O'Dea}, {Barthel}, {de
  Vries}, \& {Baum}}]{Labiano07ACS}
{Labiano}, A., {O'Dea}, C.~P., {Barthel}, P.~D., {de Vries}, W.~H., \& {Baum},
  S.~A. 2007 in prep., \,

\bibitem[{{Landolt}(1992)}]{Landolt92}
{Landolt}, A.~U. 1992, \aj, 104, 372

\bibitem[{{Large} {et~al.}(1981){Large}, {Mills}, {Little}, {Crawford}, \&
  {Sutton}}]{Large81}
{Large}, M.~I., {Mills}, B.~Y., {Little}, A.~G., {Crawford}, D.~F., \&
  {Sutton}, J.~M. 1981, \mnras, 194, 693

\bibitem[{{Marcha} {et~al.}(1996){Marcha}, {Browne}, {Impey}, \&
  {Smith}}]{Marcha96}
{Marcha}, M.~J.~M., {Browne}, I.~W.~A., {Impey}, C.~D., \& {Smith}, P.~S. 1996,
  \mnras, 281, 425

\bibitem[{{McLean} {et~al.}(1998){McLean}, {Hawkins}, {Spagna}, {Lattanzi},
  {Lasker}, {Jenkner}, \& {White}}]{McLean98}
{McLean}, B., {Hawkins}, C., {Spagna}, A., {et~al.} 1998, in IAU Symp. 179: New
  Horizons from Multi-Wavelength Sky Surveys, ed. B.~J. {McLean}, D.~A.
  {Golombek}, J.~J.~E. {Hayes}, \& H.~E. {Payne}, 431

\bibitem[{{Morganti} {et~al.}(1993){Morganti}, {Killeen}, \&
  {Tadhunter}}]{Morganti93}
{Morganti}, R., {Killeen}, N.~E.~B., \& {Tadhunter}, C.~N. 1993, \mnras, 263,
  1023

\bibitem[{{Morganti} {et~al.}(1997){Morganti}, {Tadhunter}, {Dickson}, \&
  {Shaw}}]{Morganti97}
{Morganti}, R., {Tadhunter}, C.~N., {Dickson}, R., \& {Shaw}, M. 1997, \aap,
  326, 130

\bibitem[{{O'Dea}(1998)}]{O'Dea98}
{O'Dea}, C.~P. 1998, \pasp, 110, 493

\bibitem[{{O'Dea} \& {Baum}(1997)}]{O'Dea97}
{O'Dea}, C.~P. \& {Baum}, S.~A. 1997, \aj, 113, 148

\bibitem[{{O'Dea} {et~al.}(1991){O'Dea}, {Baum}, \& {Stanghellini}}]{O'Dea91}
{O'Dea}, C.~P., {Baum}, S.~A., \& {Stanghellini}, C. 1991, \apj, 380, 66

\bibitem[{{O'Dea} {et~al.}(1996){O'Dea}, {Stanghellini}, {Baum}, \&
  {Charlot}}]{O'Dea96}
{O'Dea}, C.~P., {Stanghellini}, C., {Baum}, S.~A., \& {Charlot}, S. 1996, \apj,
  470, 806

\bibitem[{{Readhead} {et~al.}(1996){Readhead}, {Taylor}, {Pearson}, \&
  {Wilkinson}}]{Readhead96b}
{Readhead}, A.~C.~S., {Taylor}, G.~B., {Pearson}, T.~J., \& {Wilkinson}, P.~N.
  1996, \apj, 460, 634

\bibitem[{{Schlegel} {et~al.}(1998){Schlegel}, {Finkbeiner}, \&
  {Davis}}]{Schlegel98}
{Schlegel}, D.~J., {Finkbeiner}, D.~P., \& {Davis}, M. 1998, \apj, 500, 525

\bibitem[{{Sguera} {et~al.}(2005){Sguera}, {Bassani}, {Malizia}, {Dean},
  {Landi}, \& {Stephen}}]{Sguera05}
{Sguera}, V., {Bassani}, L., {Malizia}, A., {et~al.} 2005, \aap, 430, 107

\bibitem[{{Snellen} {et~al.}(1996){Snellen}, {Bremer}, {Schilizzi}, {Miley}, \&
  {van Ojik}}]{Snellen96}
{Snellen}, I.~A.~G., {Bremer}, M.~N., {Schilizzi}, R.~T., {Miley}, G.~K., \&
  {van Ojik}, R. 1996, \mnras, 279, 1294

\bibitem[{{Snellen} {et~al.}(2002){Snellen}, {Lehnert}, {Bremer}, \&
  {Schilizzi}}]{Snellen02}
{Snellen}, I.~A.~G., {Lehnert}, M.~D., {Bremer}, M.~N., \& {Schilizzi}, R.~T.
  2002, \mnras, 337, 981

\bibitem[{{Snellen} {et~al.}(2004){Snellen}, {Mack}, {Schilizzi}, \&
  {Tschager}}]{Snellen04}
{Snellen}, I.~A.~G., {Mack}, K.-H., {Schilizzi}, R.~T., \& {Tschager}, W. 2004,
  \mnras, 348, 227

\bibitem[{{Snellen} {et~al.}(1999){Snellen}, {Schilizzi}, {Bremer}, {Miley},
  {de Bruyn}, \& {R{\"o}ttgering}}]{Snellen99}
{Snellen}, I.~A.~G., {Schilizzi}, R.~T., {Bremer}, M.~N., {et~al.} 1999,
  \mnras, 307, 149

\bibitem[{{Snellen} {et~al.}(1998){Snellen}, {Schilizzi}, {de Bruyn}, {Miley},
  {Rengelink}, {R{\" o}ttgering}, \& {Bremer}}]{Snellen98}
{Snellen}, I.~A.~G., {Schilizzi}, R.~T., {de Bruyn}, A.~G., {et~al.} 1998,
  \aaps, 131, 435

\bibitem[{{Snellen} {et~al.}(2000){Snellen}, {Schilizzi}, {Miley}, {de Bruyn},
  {Bremer}, \& {R{\" o}ttgering}}]{Snellen00}
{Snellen}, I.~A.~G., {Schilizzi}, R.~T., {Miley}, G.~K., {et~al.} 2000, \mnras,
  319, 445

\bibitem[{{Spergel} {et~al.}(2003){Spergel}, {Verde}, {Peiris}, {Komatsu},
  {Nolta}, {Bennett}, {Halpern}, {Hinshaw}, {Jarosik}, {Kogut}, {Limon},
  {Meyer}, {Page}, {Tucker}, {Weiland}, {Wollack}, \& {Wright}}]{Spergel03}
{Spergel}, D.~N., {Verde}, L., {Peiris}, H.~V., {et~al.} 2003, \apjs, 148, 175

\bibitem[{{Spoelstra} {et~al.}(1985){Spoelstra}, {Patnaik}, \&
  {Gopal-Krishna}}]{Spoelstra85}
{Spoelstra}, T.~A.~T., {Patnaik}, A.~R., \& {Gopal-Krishna}. 1985, \aap, 152,
  38

\bibitem[{{Stanghellini} {et~al.}(1993){Stanghellini}, {O'Dea}, {Baum}, \&
  {Laurikainen}}]{Stanghellini93}
{Stanghellini}, C., {O'Dea}, C.~P., {Baum}, S.~A., \& {Laurikainen}, E. 1993,
  \apjs, 88, 1

\bibitem[{{Stickel} {et~al.}(1996){Stickel}, {Rieke}, {Kuehr}, \&
  {Rieke}}]{Stickel96}
{Stickel}, M., {Rieke}, G.~H., {Kuehr}, H., \& {Rieke}, M.~J. 1996, \apj, 468,
  556

\bibitem[{{Tadhunter} {et~al.}(1993){Tadhunter}, {Morganti}, {di
  Serego-Alighieri}, {Fosbury}, \& {Danziger}}]{Tadhunter93}
{Tadhunter}, C.~N., {Morganti}, R., {di Serego-Alighieri}, S., {Fosbury},
  R.~A.~E., \& {Danziger}, I.~J. 1993, \mnras, 263, 999

\bibitem[{{Tinti} {et~al.}(2005){Tinti}, {Dallacasa}, {de Zotti}, {Celotti}, \&
  {Stanghellini}}]{Tinti05b}
{Tinti}, S., {Dallacasa}, D., {de Zotti}, G., {Celotti}, A., \& {Stanghellini},
  C. 2005, \aap, 432, 31

\bibitem[{{Torniainen} {et~al.}(2005){Torniainen}, {Tornikoski},
  {Ter{\"a}sranta}, {Aller}, \& {Aller}}]{Torniainen05}
{Torniainen}, I., {Tornikoski}, M., {Ter{\"a}sranta}, H., {Aller}, M.~F., \&
  {Aller}, H.~D. 2005, \aap, 435, 839

\bibitem[{{Urry} \& {Padovani}(1995)}]{Urry95}
{Urry}, C.~M. \& {Padovani}, P. 1995, \pasp, 107, 803

\bibitem[{{White} \& {Becker}(1992)}]{White92}
{White}, R.~L. \& {Becker}, R.~H. 1992, \apjs, 79, 331

\bibitem[{{Whittle}(1985)}]{Whittle85}
{Whittle}, M. 1985, \mnras, 213, 1

\bibitem[{{Wright} \& {Otrupcek}(1990)}]{Wright90}
{Wright}, A. \& {Otrupcek}, R. 1990, in PKS Catalog (1990), 0

\bibitem[{{Wright} {et~al.}(1994){Wright}, {Griffith}, {Burke}, \&
  {Ekers}}]{Wright94}
{Wright}, A.~E., {Griffith}, M.~R., {Burke}, B.~F., \& {Ekers}, R.~D. 1994,
  \apjs, 91, 111

\bibitem[{{Xiang} {et~al.}(2006){Xiang}, {Reynolds}, {Strom}, \&
  {Dallacasa}}]{Xiang06}
{Xiang}, L., {Reynolds}, C., {Strom}, R.~G., \& {Dallacasa}, D. 2006, \aap,
  454, 729

\bibitem[{{Xiang} {et~al.}(2002){Xiang}, {Stanghellini}, {Dallacasa}, \&
  {Haiyan}}]{Xiang02}
{Xiang}, L., {Stanghellini}, C., {Dallacasa}, D., \& {Haiyan}, Z. 2002, \aap,
  385, 768

\bibitem[{{Xu} {et~al.}(1994){Xu}, {Lawrence}, {Readhead}, \& {Pearson}}]{Xu94}
{Xu}, W., {Lawrence}, C.~R., {Readhead}, A.~C.~S., \& {Pearson}, T.~J. 1994,
  \aj, 108, 395

\bibitem[{{Zensus} {et~al.}(2002){Zensus}, {Ros}, {Kellermann}, {Cohen},
  {Vermeulen}, \& {Kadler}}]{Zensus02}
{Zensus}, J.~A., {Ros}, E., {Kellermann}, K.~I., {et~al.} 2002, \aj, 124, 662

\end{thebibliography}

\begin{longtable}{ccclccccccccc}
\caption{\label{masterlist} Master list.}\\
\hline\hline
(1)     &       (2)     &       (3)     &       (4)     &       (5)     &       (6)     &       (7)     &       (8)     &       (9)     &       (10)    &       (11)    &       (12)    \\
Name & Other & GPS/HFP & ID & Redshift & Refs. & $\nu_{peak}$ & Ref. & $\nu_{peak int}$ &       Flux & Frequency & Ref.\\
        &               &               &               &       &               &       (GHz    )&              &       (GHz)   &               (Jy)    &       (GHz)   &       \\
\hline
\endfirsthead
\caption{continued.}\\
\hline\hline
(1)     &       (2)     &       (3)     &       (4)     &       (5)     &       (6)     &       (7)     &       (8)     &       (9)     &       (10)    &       (11)    &       (12)    \\
Name & Other & GPS/HFP & ID & Redshift & Refs. & $\nu_{peak}$ & Ref. & $\nu_{peak int}$ &       Flux & Frequency & Ref.\\
        &               &               &               &       &               &       (GHz    )&              &       (GHz)   &               (Jy)    &       (GHz)   &       \\
\hline
\endhead
\hline
\endfoot
\object{000319+212944} & \object{0000+212}                  &       HFP     &       G       &       0.4     &       5       &       6.2     &       6       &               &       0.265   &       5       &       6       \\
\object{000346+480703} & \object{0001+478}                  &       GPS     &               &               &       7               &       1       &       N       &       1       &       0.197   &       4.85    &       9       \\
\object{000520+052410} & \object{0002+051}                  &       HFP     &       Q       &       1.887   &       5       &       4.9     &       6       &       14.1    &       0.229   &       5       &       6       \\
\object{001052--415310} & \object{0008--42}                 &       GPS     &       G       &               &       2               &       0.5     &       N       &       0.5     &       1.12    &       5       &       10      \\
\object{002127+731241} & \object{0018+729}                  &       GPS     &       G       &       0.821   &       8,29    &       1       &       N       &       1.8     &       0.393   &       4.85    &       9       \\
\object{002225+001456} & \object{4C+00.02}       &       GPS     &       G       &       0.305   &       29              &       0.7     &       11      &       0.9     &       1.1     &       5       &       11      \\
\object{002442--420203} & \object{0022--423}                 &       GPS     &       Q       &       0.937   &       2       &               1.6     &       0       &       3.1     &       1.7     &       5       &       0       \\
\object{002914+345632} & \object{0026+346}                  &       GPS     &       G       &       0.517   &       1,12    &       1       &       N       &       1.5     &       1.32    &       4.85    &       9       \\
\object{003732+080813} & \object{0034+078}                  &       HFP     &       G?      &       $>$1.8  &       5       &       4.9     &       6       &       $>$13.7 &       0.292   &       5       &       6       \\
\object{004204+232001} & \object{0039+230}                  &       GPS     &               &               &       1               &       1?      &       N       &       1?      &       1.65    &       4.85    &       9       \\
\object{010813--120050} & \object{0105--122}                 &       GPS     &       G       &               &       4               &       1       &       4       &       1       &       0.52    &       2.7     &       4       \\
\object{011137+390628} & \object{0108+388}                  &       HFP     &       G       &       0.66847 &       1,29    &       4       &       11      &       6.7     &       1.26    &       5       &       11      \\
\object{011638+242253} & \object{0113+241}                  &       GPS     &               &               &       5       &       4.9     &       6       &       4.9     &       0.243   &       5       &       6       \\
\object{011935+321050} & \object{4C+31.04}		&      GPS     &       G       &       0.06    &       7,13    &       0.4     &       N       &       0.4     &       1.59    &       4.85    &       9       \\
\object{014658+211024} & \object{0144+209}                  &       GPS     &               &               &       1       &       1.3?    &       N       &       1.3?    &       0.598   &       4.85    &       9       \\
\object{015310--331025} & \object{0150--334}                 &       GPS     &       Q       &       0.61    &       3       &       1.5     &       3       &       2.4     &       0.88    &       4.8     &       3       \\
\object{020434+090349} & \object{0201+088}                  &       GPS     &               &               &       7               &       $\sim$2 &       N       &       $\sim$2 &       0.774   &       4.85    &       9       \\
\object{020346+113445} & \object{0201+113}                  &       HFP     &       Q       &       3.639   &       1,29    &       $\sim$4 &       N       &       17.3    &       0.742   &       4.85    &       9       \\
\object{020643--302458} & \object{0204--306}                 &       GPS     &       G       &               &       4               &       0.5     &       4       &       0.5     &       0.58    &       2.7     &       4       \\
\object{021010--221336} & \object{0207--224}                 &       GPS     &       G       &               &       4               &       1.5     &       4       &       1.5     &       0.85    &       2.7     &       4       \\
\object{021044+041934} & \object{0208+040}                  &       GPS     &               &               &       4               &       0.4     &       4       &       0.4     &       0.56    &       2.7     &       4       \\
\object{024008--230915} & \object{0237--233}                 &       GPS     &       Q       &       2.223   &       11,1    &       1       &       11      &       3.2     &       3.34    &       5       &       11      \\
\object{024235--213226} & \object{0240--217}                 &       GPS     &       G       &       0.314   &       4       &       1       &       4       &       1.3     &       0.97    &       2.7     &       4       \\
\object{025134+431515} & \object{0248+430}                  &       HFP     &       Q       &       1.32            &       1,29    &       7       &       29      &       16.2    &       1.24    &       5       &       30      \\
\object{031857+162833} & \object{4C+16.09}&      GPS     &       Q       &               &       11              &       0.8     &       11      &       0.8     &       2.89    &       5       &       11      \\
\object{032153+122113} & \object{0319+121}                  &       GPS     &       Q       &       2.662   &       1       &       0.4     &       11      &       1.5     &       1.1     &       5       &       11      \\
\object{032320+053411} & \object{4C+05.14}&      GPS     &       G       &       0.1785  &       4       ,14     &       0.4     &       4       &       0.5     &       1.6     &       2.7     &       4       \\
\object{035721+231953} & \object{0354+231}                  &       HFP     &       Q       &               &       5       &       $>$22   &       6       &               &       0.56    &       5       &       6       \\
\object{040121--292126} & \object{0359--294}                 &       GPS     &       G       &               &       4               &       0.4     &       4       &       0.4     &       0.58    &       2.7     &       4       \\
\object{041046+765645} & \object{4C+76.03}&      GPS     &       G       &       0.5985  &       1     &       0.6     &       0       &       1       &       2.82    &       5       &       0       \\
\object{040757--275705} & \object{0405--280}                 &       GPS     &       G       &               &       4               &       1.5     &       4       &       1.5     &       0.93    &       2.7     &       4       \\
\object{040734--392447} & \object{0405--395}                 &       GPS     &       G       &               &       4               &       0.4     &       4       &       0.4     &       0.52    &       2.7     &       4       \\
\object{042214--384452} & \object{0420--388}&       GPS     &       Q       &       3.11    &       1     &       ?       &       N       &       ?       &       0.13    &       4.85    &       26      \\
\object{042746+413301} & \object{0424+414}                  &       GPS     &               &               &       7,              &       $\sim$2 &       N       &       $\sim$2 &       0.723   &       4.85    &       9       \\
\object{043103+203734} & \object{0428+205}			& GPS     &       G       &       0.219   &       1,29    &       1.1     &       11      &       1.3     &       2.38    &       5       &       11      \\
\object{043354--022956} & \object{4C --02.17}			&    GPS     &       G       &               &       4       &       0.4     &       4       &       0.4     &       1.04    &       2.7     &       4       \\
\object{043701--184448} & \object{0434--188}                 &       HFP     &       Q       &       2.702   &       3       &       4.5     &       3       &       16.7    &       0.95    &       4.8     &       3       \\
\object{044133--334003} & \object{0439--337}                 &       GPS     &               &               &       4               &       1.5     &       4       &       1.5     &       0.88    &       2.7     &       4       \\
\object{045720--084905} & \object{0454--088}                 &       GPS     &       G       &               &       4               &       0.4     &       4       &       0.4     &       0.58    &       2.7     &       4       \\
\object{045952+022931} & \object{0457+024}                  &       HFP     &       Q       &       2.384   &       1,29    &       2.1     &       11      &       7.1     &       1.57    &       5       &       11      \\
\object{050321+020305} & \object{0500+019}		&   GPS     &       Q       &       0.58457 &       2,7     &       1.8     &       11      &       2.9     &       1.89    &       5       &       11      \\
\object{051002+180042} & \object{0507+179}                  &       GPS     &       Q       &       0.3     &       29      &       1.4     &       0       &       1.8     &       0.73    &       5       &       0       \\
\object{053008--250330} & \object{0528--250}                 &       HFP     &       Q       &       2.813   &       1,29    &       2.7     &       N       &       10.3    &       1.16    &       5       &       10      \\
\object{055652--024105} & \object{0554--026}                 &       GPS     &       G       &       0.235   &       18              &       1       &       0       &       1.2     &       0.29    &       5       &       0       \\
%\object{0615+820} &               &       HFP     &       Q       &       0.71    &       1       &       $\sim$4.5       &       N       &       7.7     &       0.96    &       2.3     &       10      \\
\object{062518+444002} & \object{0621+446}                  &       HFP     &               &               &       5               &       14      &       6       &               &       0.442   &       5       &       6       \\
\object{063802+593322} & \object{0633+595}                  &       HFP     &               &               &       5               &       12.9    &       6       &               &       0.591   &       5       &       6       \\
\object{064204+675836} & \object{0636+680}                  &       HFP     &       Q       &       3.18    &       1       &       3.7     &       6       &       15.5    &       0.474   &       5       &       6       \\
\object{064425--345942} & \object{0642--349}                 &       HFP     &       Q       &       2.165   &       3       &       3.3     &       3       &       10.5    &       0.85    &       4.8     &       3       \\
\object{064632+445117} & \object{0642+449}                  &       HFP     &       Q       &       3.396   &       5       &       15.5    &       6       &       68.1    &       1.896   &       5       &       6       \\
\object{065031+600143} & \object{0646+600}                  &       HFP     &       Q       &       0.455   &       1,5     &       6.8     &       6       &       9.9     &       1.236   &       5       &       6       \\
\object{070648+464756} & \object{0703+468}                  &       GPS     &       Q?      &               &       7               &       0.5     &       0       &       0.5     &       0.62    &       4.85    &       9       \\
\object{071338+434917} & \object{0710+439}                  &       GPS     &       G       &       0.518   &       1,11    &       1.9     &       11      &       2.9     &       1.68    &       5       &       11      \\
\object{071424+353439} & \object{0711+356}                  &       GPS     &       Q       &       1.62    &       1,29    &       1.4     &       N       &       3.7     &       0.89    &       4.85    &       9       \\
\object{071509+452555}& \object{0711+453}					  &       GPS     &       G       &       0.042   &       16      &       3.8     &       16      &       4.0     &       0.074   &       1.4     &       16      \\
\object{072550+391725} &\object{4C+39.17}&      GPS     &       G       &               &       7       &       0.5?    &       N       &       0.5?    &       0.23    &       4.85    &       9       \\
\object{073328+560541}& \object{0729+562}  &       GPS     &       G       &       0.104   &       16      &       0.46    &       16      &       0.5     &       0.394   &       1.4     &       16      \\
\object{073934+495438}& \object{0735+500}  &       GPS     &       G       &       0.054   &       16      &       0.95    &       16      &       1       &       0.107   &       1.4     &       16      \\
\object{074533+101112} & \object{0742+103}                  &       HFP     &       G       &       2.624   &       17              &       2.7     &       11      &       9.8     &       3.46    &       5       &       11      \\
\object{074554--004418} & \object{0743--006}                 &       HFP     &       Q       &       0.994   &       17              &       5.8     &       11      &       11.6    &       2.05    &       5       &       11      \\
\object{075415+532456} & \object{0750+535}                  &       GPS     &               &               &       7               &       1.4     &       N       &       1.4     &       0.29    &       4.85    &       9       \\
\object{080454+433537}& \object{0801+437 } &        GPS     &       Q       &       0.123   &       16      &       1.5     &       16      &       1.7     &       0.36    &       1.4     &       16      \\
\object{080538+210651} & \object{0802+212}                   &       GPS     &       G       &               &       1               &       1.4     &       N       &       1.4     &       0.56    &       4.85    &       9       \\
\object{083139+460800} & \object{0828+461}  &       GPS     &       G       &       0.127   &       16      &       2.2     &       16      &       2.5     &       0.131   &       1.4     &       16      \\
\object{090040--280820} & \object{0858--279}                 &       GPS     &       Q       &       2.16            &       1,29    &       1.4     &       29      &       4.4     &       1.38            &       5       &       27      \\
\object{090615+463618} & \object{0902+468}        &        GPS     &       G       &       0.085   &       16      &       0.68    &       16      &       0.7     &       0.314   &       1.4     &       16      \\
\object{090641+034242} & \object{0904+039}                  &       GPS     &       G       &               &       1               &       $\sim$0.6       &       N       &       $\sim$0.6       &       0.208   &       4.85    &       9       \\
\object{091335+145420} & \object{0910+151}                  &       GPS     &               &               &       4               &       0.6     &       4       &       0.6     &       0.54    &       2.7     &       4       \\
\object{091716+111336} & \object{0914+114}                  &       GPS     &               &               &       11              &       0.4?    &       N       &       0.4?    &       0.13    &       5       &       0       \\
\object{093609+331308}& \object{0933+332}  &       GPS     &       G       &       0.076   &       16      &       2.2     &       16      &       2.4     &       0.055   &       1.4     &       16      \\
\object{094336--081931} & \object{0941--080}                 &       GPS     &       G       &       0.228   &       11,1    &       0.5     &       11      &       0.6     &       1.11    &       5       &       11      \\
\object{103507+562847} & \object{1031+567}                  &       GPS     &       Q       &       0.459   &       29      &       1.3     &       11      &       1.9     &       1.28    &       5       &       11      \\
\object{104437--271218} & \object{1042--269}                 &       GPS     &               &               &       4               &       1.5     &       4       &       1.5     &       0.55    &       2.7     &       4       \\
\object{105715+001203} & \object{1054+004}                  &       GPS     &               &               &       4               &       0.4     &       4       &       0.4     &       0.58    &       2.7     &       4       \\
\object{105731+405646} & \object{NGC 3468}        &       GPS     &       G       &       0.008   &       16      &       1.25    &       16      &       1.3     &       0.047   &       1.4     &       16      \\
\object{110323+220337} & \object{1100+223}                  &       GPS     &               &               &       1       &       2.7     &       N       &       2.7     &       0.58    &       4.85    &       9       \\
\object{110946+104343} & \object{1107+109}                  &       GPS     &       G       &               &       4               &       0.5     &       4       &       0.5     &       0.8     &       2.7     &       4       \\
\object{111000--185848} & \object{1107--187}                 &       GPS     &       G       &       0.497   &       4       &       1       &       4       &       1.5     &       0.65    &       2.7     &       4       \\
\object{111120+195536} & \object{1108+201}                  &       GPS     &       G       &       0.299   &       7               &       1       &       N       &       1.3     &       0.64    &       4.85    &       9       \\
\object{112027+142054} & \object{4C+14.41}&      GPS     &       G       &       0.362   &       11      &       0.5     &       11      &       0.7     &       1       &       5       &       11      \\
\object{112125--055356} & \object{1118--056}                 &       GPS     &               &               &       29              &       0.9     &       29      &               &       0.57    &       5       &       27 \\
\object{112256--274248} & \object{1120--274}                 &       GPS     &               &               &       4               &       1.4     &       4       &       1.4     &       0.74    &       2.7     &       4       \\
\object{113007--144927} & \object{1127--145}                 &       GPS     &       Q       &       1.187   &       11      &       1       &       11      &       2.2     &       3.82    &       5       &       11      \\
\object{113513--002119} & \object{4C--00.45}&     GPS     &       Q       &               &       4               &       0.4     &       4       &       0.4     &       0.76    &       2.7     &       4       \\
\object{113555+425844} & \object{1133+432}                  &       GPS     &               &               &       15      &       1.0     &       29      &       $\sim$1 &       0.42    &       5       &       15      \\
\object{114608--244731} & \object{1143--245}                 &       GPS     &       Q       &       1.95    &       11      &       2.2     &       11      &       6.5     &       1.4     &       5       &       11      \\
\object{120321+041417} & \object{1200+045}                  &       GPS     &       G       &       1.21177 &       4               &       0.4     &       4       &       0.9     &       0.52    &       2.7     &       4       \\
\object{122758+363511} & \object{1225+36}                   &       GPS     &       Q       &       1.973   &       1,24    &       1.2     &       11      &       3.6     &       0.77    &       5       &       11      \\
\object{124823--195918} & \object{1245--197}                 &       GPS     &       Q       &       1.275   &       1,29    &       0.5     &       11      &       1.1     &       2.34    &       5       &       11      \\
\object{130041--105908} &\object{1258-104} &       GPS     &       Q?      &       1.283?  &       19.28   &       ?       &       N       &       ?       &       0.07    &       4.85    &       25      \\ %\object{PMN J1300--1059}
\object{131338+693909} &   \object{1312+695}           &       GPS     &               &               &       1               &       ?       &       N       &       ?       &       0.26    &       4.85    &       9       \\
\object{131739+411545} & \object{1315+415} &        GPS     &       G       &       0.066   &       16      &       2.3     &       16      &       2.5     &       0.249   &       1.4     &       16      \\
\object{132616+315409} & \object{4C+32.44}&      GPS     &       G       &       0.369   &       29      &       0.5     &       11      &       0.7     &       2.39    &       5       &       11      \\
\object{132513+395552}    &\object{1322+401} &       GPS     &       G       &       0.074   &       16      &       1.9     &       16      &       2       &       0.056   &       1.4     &       16      \\
\object{133522+454238} & \object{1333+459}                  &       HFP     &               &       2.45    &       29,5    &       5       &       29      &       17      &       0.79    &       5       &       5       \\
\object{133525+584400} & \object{4C +58.26}                  &       GPS     &               &               &       5       &       4.9     &       6       &       4.9     &       0.723   &       5       &       6       \\
\object{134551--301504} & \object{1343--300}                 &       GPS     &               &               &       4               &       0.4     &       4       &       0.4     &       0.56    &       2.7     &       4       \\
\object{134035+444817}& \object{1338+450}   &       GPS     &       G       &       0.065   & 16    &       2.3     &       16      &       2.4     &       0.082   &       1.4     &       16      \\
\object{134733+121724} & \object{4C+12.50}&      GPS     &       G       &       0.12174 &       11,1    &       0.4     &       11      &       0.4     &       3.05    &       5       &       11      \\
\object{135014--220441} & \object{1347--218}                 &       GPS     &       G       &               &       4               &       0.4     &       4       &       0.4     &       0.72    &       2.7     &       4       \\
\object{135230+023247} & \object{1349+027}                  &       GPS     &       Q       &               &       4               &       0.4     &       4       &       0.4     &       0.78    &       2.7     &       4       \\
\object{135256+110707} & \object{4C+11.46}&      GPS     &               & &     4               &       0.4     &       4       &       0.4     &       1.04    &       2.7     &       4       \\
%\object{1351--018}        &               &       GPS     &               &       3.707   &       1       &       0.4     &       4       &       1.9     &       0.78    &       2.7     &       4       \\
\object{135706--174402} & \object{1354--174}                 &       GPS     &       Q       &       3.147   &       1,29    &       1.2     &       N       &       5       &       0.97    &       5       &       27      \\
\object{140028+621038} & \object{4C+62.22}&      GPS     &       G       &       0.431   &       11,7    &       0.5     &       11      &       0.7     &       1.8     &       5       &       11      \\
\object{140700+282714} & \object{OQ208}& GPS     &       G       &       0.0769  &       11,5    &       4.2     &       11      &       4.5     &       2.69    &       5       &       11      \\
\object{141236+133438} & \object{1410+138}                  &       GPS     &               &               &       5       &       4.2     &       6       &       4.2     &       0.33    &       5       &       6       \\
%\object{1413+349} &\object{OQ 323}&       GPS     &               &               &       1       &       1       &       11      &       1       &       1.02    &       5       &       11      \\
\object{142438+225601} & \object{1422+231}                  &       HFP     &       Q       &       3.626   &       5               &       4.0     &       6       &       18.5    &       0.61    &       5       &       6       \\
\object{143009+104328} & \object{1427+109}                  &       HFP     &       Q       &       1.71    &       5               &       4.9     &       6       &       13.3    &       0.91    &       5       &       6       \\
\object{143539--041455} & \object{1433--04}                 &       GPS     &       G       &       0.795   &       11      &       0.6     &       N       &       1.1     &       0.2     &       5       &       0       \\
\object{144516+095836} & \object{OQ172}& GPS     &       Q       &       3.535   &       11      &       0.9     &       11      &       4.1     &       1.2     &       5       &       11      \\
\object{144516+095836} & \object{1444--339}                 &       GPS     &       G       &               &       4               &       0.5     &       4       &       0.5     &       0.5     &       2.7     &       4       \\
\object{150506+032630} & \object{1502+036}                  &       HFP     &       Q       &       0.411   &       5       &       6.2     &       6       &       8.8     &       0.93    &       5       & 6       \\
\object{150603--091912} & \object{1503--091}                 &       GPS     &       G       &               &       4               &       0.6     &       4       &       0.6     &       0.87    &       2.7     &       4       \\
\object{151141+051809} & \object{1509+054}                  &       HFP     &       G       &       0.084   &       5               &       11.0    &       6       &       11.9    &       0.54    &       5       &       6       \\
\object{152114+043022} & \object{4C+04.51}&     GPS     &       Q       &       1.296   &       29      &       0.8     &       11      &       1.8     &       1.09    &       5       &       11      \\
\object{152237--273010} & \object{1519--273}                 &       HFP     &       Q       &       1.294   &       29,3    &       5.8     &       3       &       6.9     &       1.74    &       4.8     &       3       \\
\object{152642+665054} & \object{1526+670}                  &       HFP     &       Q       &       3.02    &       5       &       5.8     &       6       &       23.3    &       0.41    &       5       &       6       \\
\object{154301--075707} & \object{1540--077}                 &       GPS     &       G       &       0.172   &       4       &       0.4     &       4       &       0.5     &       1.21    &       2.7     &       4       \\
\object{154609+002624} & \object{1543+005}                  &       GPS     &       G       &       0.556   &       18,4    &       1.2     &       0       &       1.9     &       0.84    &       5       &       0       \\
\object{154812--121331} & \object{1545--120}                 &       GPS     &       G       &       0.883   &       4       &       0.4     &       4       &       0.8     &       1.45    &       2.7     &       4       \\
\object{155614--062235} & \object{4C--06.43}&     GPS     &       G       &               &       4 &     0.4     &       4       &       0.4     &       0.77    &       2.7     &       4       \\
\object{160000--003723} & \object{1557--004}                 &       GPS     &               &        &      4                       &       1       &       4       &       1       &       0.54    &       2.7     &       4       \\
\object{160207+332653} & \object{1600+335}		& GPS     &       G       &       1.1     &       11,23   &       2.4     &       11      &       5       &       2.67    &       5       &       11      \\
%\object{1601--222}        &               &       GPS     &       G       &       0.141   &       1,4     &       0.6     &       4       &       0.7     &       0.57    &       2.7     &       4       \\
\object{160631+312710} & \object{1604+315}                  &       GPS     &       G       &       1.5p    &       1,31    &       &       1.5     &       29      &       0.08    &       4.8     &       9       \\
\object{160913+264129} & \object{CTD93}& GPS     &       G       &       0.473   &       1,29    &       1.1     &       11      &       1.6     &       1.73    &       5       &       11      \\
\object{161637+045932} & \object{1614+051}                  &       HFP     &       Q       &       3.197&  1,6     &       4.1     &       6       &       17.2    &       0.89    &       5       &       6       \\
\object{162418--680913} & \object{1619--680}                 &       GPS     &       Q       &       1.36    &       3               &       3.1     &       3       &       7.3     &       1.69    &       4.8     &       3       \\
\object{162304+662401} & \object{1622+665}                  &       HFP     &       G       &       0.203   &       5               &       5.1     &       6       &       6.1     &       0.3     &       5       &       6       \\
\object{164047+122002} & \object{4C+12.6}&       GPS     &       G       &       1.152   &       4       &       0.4     &       4       &       0.9     &       1.48    &       2.7     &       4       \\
\object{164558+633011} & \object{1645+635}                  &       HFP     &       Q       &       2.379   &       5       &       $>$22   &       6       &       $>$74.3 &       0.51    &       5       &       6       \\
\object{164831+024248} & \object{4C+02.43}&      GPS     &               &       &4              &       0.4     &       4       &       0.4     &       0.61    &       2.7     &       4       \\
\object{165103+012923} & \object{1648+015}                  &       GPS     &       Q       &       0.4     &       0               &       1.5     &       0       &       2.1     &       1.03    &       5       &       0       \\
\object{165844--073918} & \object{1656--075}                 &       GPS     &               &               &       3       &       4.8     &       3       &       4.8     &       1.32    &       4.8     &       3       \\
\object{172340--650036} & \object{1718--649}                 &       GPS     &       G       &       0.014   &       0       &       4       &       0       &       4.1     &       4.32    &       5       &       0       \\
\object{172657--642753} & \object{1722--644}                 &       GPS     &               &               &       3       &       1.1     &       3       &       1.1     &       1.26    &       4.8     &       3       \\
\object{173458+092657} & \object{1732+094}                  &       GPS     &       G       &       0.61p   &       7,4     &       2.8     &       0       &       4.5     &       0.86    &       5       &       0       \\
\object{173549+504911} & \object{1734+508}                  &       HFP     &       G?      &               &       5       &       5.9     &       6       &               &       0.97    &       5       &       6       \\
\object{174425--514444} & \object{1740--517}                 &       GPS     &       G       &               &       2               &       1?      &       N       &       1?      &       3.9     &       5       &       27      \\
\object{175301+275059} & \object{1751+278}                  &       GPS     &       G       &       0.86p   &       1,15    &       0.66?   &       N       &       1.2     &       0.27    &       5       &       15      \\
\object{171854+544148}& \object{1753+544}  &       GPS     &               &       0.147   &       16      &       0.48    &       16      &       0.6     &       0.33    &       1.4     &       16      \\
\object{180356+034108} & \object{1801+036}                  &       GPS     &       G       &               &       7       &       ?       &       N       &       ?       &       0.25    &       4.85    &       9       \\
\object{181944+670847} & \object{1819+671}                  &       GPS     &       G       &       0.22    &       7,21    &       $\sim$0.5       &       N       &       0.6     &       0.15    &       4.85    &       9       \\
\object{182632+270808} & \object{1824+271}                  &       GPS     &       G?      &               &       15,29   &       1?      &       N       &       1?      &       0.115   &       5       &       15      \\
\object{183728--710844} & \object{1831--711}                 &       HFP     &       Q       &       1.356   &       3       &       8.2     &       3       &       19.3    &       2.39    &       4.8     &       3       \\
\object{184057+390046} & \object{1839+389}                  &       HFP     &       Q       &       3.095   &       5               &       4.5     &       6       &       18.4    &       0.203   &       5       &       6       \\
\object{184103+671849} & \object{1841+673}                  &       GPS     &       G       &       0.47    &       0               &       2?      &       N       &       2.9?    &       0.16    &       5       &       9       \\
\object{184535+354116} & \object{1843+356}                  &       GPS     &       Q       &       0.764   &       1,7     &       2       &       0       &       3.5     &       0.82    &       5       &       0       \\
\object{185027+282514} & \object{1848+283}                  &       HFP     &       Q       &       2.56    &       1,5     &       8.3     &       6       &       29.5    &       1.246   &       5       &       6       \\
\object{185527+374257} & \object{1853+376}                  &       HFP     &       G       &       0.5     &       5               &       4.5     &       6       &       6.8     &       0.36    &       5       &       6       \\
\object{193925--634246} & \object{1934--638}                 &       GPS     &       G       &       0.183   &       2,1     &       1.4     &       0       &       1.7     &       6.5     &       5       &       0       \\
\object{194553+705550} & \object{1946+708}                  &       GPS     &       G       &       0.101   &       7,21    &       $\sim$0.6       &       N       &       0.7     &       0.64    &       4.85    &       9       \\
\object{200324--325147} & \object{2000--330}                 &       HFP     &       Q       &       3.773   &       1       &       5       &       N       &       23.9    &       1.2     &       5       &       27      \\
\object{201114--064403} & \object{2008--068}                 &       GPS     &       G       &       0.547   &       4,29            &       1.4     &       11      &       2.2     &       1.34    &       5       &       11      \\
\object{202135+051505} & \object{2019+050}                  &       GPS     &       Q       &               &       5       &       3.7     &       6       &       3.7     &       0.477   &       5       &       6       \\
\object{202456+171814} & \object{2022+171}                  &       HFP     &       G       &       0.9     &       5               &       14.5    &       6       &               &       0.57    &       5       &       6       \\
\object{205252+363535} & \object{2050+364}                  &       GPS     &       G       &       0.354   &       1,22    &       1.2     &       N       &       1.6     &       3.4     &       4.85    &       9       \\
\object{205828+054251} & \object{4C+05.78}&      GPS     &       G       &       1.381   &       4       &       0.4     &       4       &       1       &       0.65    &       2.7     &       4       \\
\object{212339--011234} & \object{2121--014}                 &       GPS     &       Q       &       1.158   &       20      &       0.5     &       0       &       1.1     &       0.32    &       5       &       0       \\
\object{212912--153841} & \object{2126--158}                 &       GPS     &       Q       &       3.27    &       1,11    &       4.1     &       11      &       17.5    &       1.17    &       5       &       11      \\
\object{213032+050217} & \object{2128+048}		&        GPS     &       G       &       0.99    &       11,4    &       0.7     &       11      &       1.4     &       2.02    &       5       &       11      \\
\object{215203--780707} & \object{2146--783}                 &       GPS     &       Q       &               &       3       &       4.3     &       3       &       4.3     &       1.15    &       4.8     &       3       \\
\object{215137+055213} & \object{2149+056}                  &       GPS     &       G       & 0.74  &       29      &       4.0     &       29      &       5.9     &       1.19    &       5       &       27\\
\object{215550--113948} &    \object{2153--115}     &       GPS     &               &               &       1               &       1?      &       N       &       1?      &       0.37    &       4.85    &       25      \\
\object{221206+235540} & \object{2209+236}                  &       HFP     &       Q       &               &       5       &       12.6    &       6       &               &       1.18    &       5       &       6       \\
\object{221237+015251} & \object{4C+01.69}&      GPS     &       G       &       &11             &       0.5     &       11      &       0.5     &       1.05    &       5       &       11      \\
\object{223834+124251} & \object{2236+124}                  &       GPS     &       Q       &               &       1,29    &       5?      &       N       &       5?      &       0.33    &       5       &       27      \\
\object{225717+024317} & \object{2254+024}  &    HFP     &       Q       &       2.081   &       5       &       19.5    &       6       &       60.1    &       0.274   &       5       &       6       \\
\object{232510--034446} & \object{2322--040}                 &       GPS     &       G       &               &       1,29            &       1.4     &       4       &       1.4     &       0.91    &       2.7     &       4       \\
\object{232503+791716} & \object{2323+790}                 &       GPS     &       G       &               &       1,29            &       ?       &       N       &       ?       &       1.136   &       1.4     &       19      \\
\object{233013+334838} & \object{2327+335}     &       HFP     &       Q       &       1.809   &       5               &       5.6     &       6       &       15.7    &       0.558   &       5       &       6       \\
\object{233946--060412} & \object{4C--06.76}&     GPS     &       G       & &     4       &       0.4     &       4       &       0.4     &       0.8     &       2.7     &       4       \\
\object{234029+264157} & \object{2337+264}                  &       GPS     &       Q       &               &       1,29            &               &               &               &               &               &               \\
\object{234403+822640} & \object{2342+821}                  &       GPS     &       Q       &       0.735   &       11,1    &       0.5     &       11      &       0.9     &       1.28    &       5       &       11      \\
%\object{2352+495}      &               &       GPS     &       G       &       0.237   &       11,7    &       0.7     &       11      &       0.9     &       1.49    &       5       &       11      \\
\end{longtable}

Columns: 
(1) J2000 name
(2) B1950 or catalogue name.%, when available
%(2) Other name: catalogue name, or name as listed in the reference paper.
(3) GPS/HFP classification according to intrinsic peak frequency.
(4) Optical identification: G = galaxy, Q = quasi-stellar object.
(5) Redshift (p = photometric).
(6) Reference where the source is listed as a GPS, and references for optical ID and redshift. 
(7 and 8) Observed frequency of the spectral peak and reference. Data from NED were used to estimate the spectral peak for those sources with no published measurements of the peak frequency.  "?" means that confirmation of the radio spectral shape seems necessary.%Spectral peaks form NED are estimations..
(9) Intrinsic frequency of the spectral peak.
(10, 11 and 12) Flux density, frequency at which the flux density was measured, and reference.\\

References:
0 = This work.
1 = \citet{O'Dea91},
2 = 2 Jy Sample: \citet{Morganti93, Tadhunter93,Serego94, Morganti97},
3 = \citet{Edwards04},
4 = \citet{Snellen02},
5 = \citet{Tinti05b},
6 = \citet{Dallacasa00},
7 = \citet{Augusto06},
8 = \citet{Snellen96},
9 = \citet{Becker91},
10 = \citet{Kuehr81},
11 = \citet{O'Dea98},
12 = \citet{Zensus02},
13 = \citet{Marcha96},
14 = \citet{Best03},
15 = \citet{Xiang06},
16 = CORALZ sample: \citet{Snellen04},
17 = \citet{Torniainen05},
18 = \citet{Vries00b},
19 = \citet{Condon98},
20 = \citet{Vries95},
21 = \citet{Snellen99},
22 = \citet{Beasley02},
23 = \citet{Snellen00},
24 = \citet{Xu94},
25 = \citet{Griffith94},
26 = \citet{Wright94},
27 = \citet{Wright90},
28 = \citet{Breuck02},
29 = \citet{Vries97b},
30 = \citet{Snellen98},
31 = \citet{Xiang02},
N = NASA/IPAC Extragalactic Database (NED).

\end{document}